\documentclass[11pt,a4paper]{article}
\RequirePackage{etex} 
\usepackage{etex}

\usepackage{multirow}
\usepackage[abs]{overpic}
\usepackage{graphicx}
\usepackage{pgfplots}
\usepackage{pgfplotstable} 
\usepgfplotslibrary{statistics}
\usepackage{subfigure}
\usetikzlibrary{arrows,decorations.pathmorphing,backgrounds,positioning,fit,calc,3d,er,pgfplots.groupplots}
\usetikzlibrary{spy}
\usepgfplotslibrary{fillbetween}
\usetikzlibrary{patterns}

\usepackage{booktabs}
\usepackage{comment}

\pgfplotsset{compat=1.16}

\usepackage[bold]{hhtensor}
\usepackage{float}
\usepackage{tikz}
\usetikzlibrary{positioning,fit,calc,decorations.pathreplacing}

\definecolor{diffcol}{RGB}{252,244,214}
\definecolor{inscol}{RGB}{221,238,255}
\definecolor{sinkcol}{RGB}{214,237,214}
\definecolor{reaccol}{RGB}{245,214,214}
\definecolor{doseColor}{HTML}{0072B2} 
\definecolor{tempColor}{HTML}{D55E00} 
\definecolor{cwColor}{HTML}{009E73}   
\definecolor{reColor}{HTML}{CC79A7}   

\usepackage{cases}
\usetikzlibrary{arrows,decorations.pathmorphing,backgrounds,positioning,fit,calc,3d,er}
\usetikzlibrary{intersections,fillbetween}

\usepackage[most]{tcolorbox}
\usepackage{soul}

\usepackage{hyperref}
\hypersetup{
    colorlinks,%
    citecolor=blue,%
    filecolor=black,%
    linkcolor=magenta,%
    urlcolor=black
}
\def\BarWi{5pt}        
\def\ShiftA{-18pt}      
\def\ShiftB{0pt}        
\def\ShiftC{25pt}  

\usepackage{latexsym}
\usepgfplotslibrary{fillbetween}

\usepackage{mathtools}
\usepackage{comment}
\usepackage[usenames, dvipsnames]{xcolor}
\usepackage[margin=1in]{geometry}

\newcommand{\etal}{{\it et al.}}
\usepackage{tikz}
\usepackage{tikz-qtree}
\usetikzlibrary{shapes.geometric, arrows}
\usetikzlibrary{calc}
\tikzstyle{startstop} = [rectangle, rounded corners, minimum width=3cm, minimum height=1cm, text centered, draw=black, fill=white!30]
\tikzstyle{process} = [rectangle, minimum width=3cm, minimum height=1cm, text centered, draw=black, fill=white!30]
\tikzstyle{decision} = [diamond, minimum width=3cm, minimum height=1cm, text centered, draw=black, fill=white!30]
\tikzstyle{arrow} = [thick,->,>=stealth]
\usepackage{tkz-berge}
\usepackage{transparent}
\usepackage{sansmath}
\usepackage{bm}
\usetikzlibrary{shadings,intersections,fadings}
\usetikzlibrary{shapes.geometric, arrows.meta}
\usepackage{tikz-3dplot}
\usetikzlibrary{calc,3d,decorations.markings,backgrounds,positioning,intersections,shapes,quotes,angles}
\usepackage{nicefrac}
\usepackage{caption}
\usepackage{pdflscape}
\usepackage{caption,booktabs}
\usepackage{amsmath}
\usepackage{amstext} 

\usepackage[abs]{overpic}
\usepackage{array}
\usepackage{enumitem}

\hypersetup{
    colorlinks,%
    citecolor=blue,%
    filecolor=black,%
    linkcolor=blue,%
    urlcolor=black
}
\usepackage{cleveref}
\usepackage{tabularx}
\usepackage{multicol}
\usepackage{authblk}

\usepackage{algorithm}
\usepackage{algpseudocode}

\usepackage{scrextend}
\usepackage{lineno}
\modulolinenumbers[1]

\title{Irradiation-Driven Recrystallization in Fusion-Grade Tungsten: A Mesoscale, Microstructure-Aware Model}

\author[1]{Jinxin Yu \thanks{jinxinyu@g.ucla.edu}}
\author[1]{Sicong He}
\author[2]{Giacomo Po}
\author[3,4,5]{Jason R. Trelewicz}
\author[6]{Timothy J. Rupert}
\author[1,7]{Jaime Marian \thanks{jmarian@ucla.edu}}

\affil[1]{Department of Materials Science and Engineering, University of California Los Angeles, Los Angeles, CA 90095, USA}
\affil[2]{Department of Mechanical and Aerospace Engineering, University of Miami, Coral Gables, FL 33146, USA}
\affil[3]{Department of Materials Science and Chemical Engineering, Stony Brook University, Stony Brook, NY 11794, USA}
\affil[4]{Institute for Advanced Computational Science, Stony Brook University, Stony Brook, NY 11794, USA}
\affil[5]{Materials Science and Technology Division, Oak Ridge National Laboratory, Oak Ridge, TN 37831, USA}
\affil[6]{Hopkins Extreme Materials Institute, Johns Hopkins University, Baltimore, MD 21218, USA}
\affil[7]{Department of Mechanical and Aerospace Engineering, University of California, Los Angeles, CA 90095, USA}

\begin{document}
\maketitle
\begin{abstract}
Tungsten (W) is the leading candidate material for plasma-facing components in fusion reactors, yet its upper operational temperature is limited by premature grain growth and recrystallization processes. Irradiation adds further complications by generating defect clusters and transmutation products that alter both the driving forces and kinetics of grain boundary motion. Designing W alloys with better high-temperature resistance is a necessity for the further development of advanced fusion energy devices. However, the complexities associated with all the physical processes that take part in recrystallization under irradiation, plus the poor general reproducibility of W microstructures with sufficient fracture toughness, makes this task an extremely challenging undertaking.
In this work, we develop a physics-based, multiscale framework that couples crystal plasticity, stochastic cluster dynamics, and discrete grain boundary dynamics to model the co-evolution of plastic deformation, irradiation damage, and grain growth in fusion-grade tungsten polycrystals. The approach enables simulations on realistic microstructures with arbitrary grain size and misorientation distributions, without recourse to mean-field simplifications.
The model captures (i) the spatial heterogeneity of dislocation density distribution during hot working; (ii) irradiation-induced defect accumulation under fusion conditions, and (iii) the buildup of chemical and elastic driving forces for grain boundary migration and microstructural evolution. Our simulations reveal an incubation period followed by rapid irradiation-assisted recrystallization characterized by Avrami exponents close to unity, which is consistent with experimental observations. Parametric studies demonstrate the dominant influence that temperature has on thermally activated grain-boundary mobility, a weaker dependence on prior strain, and a pronounced retardation of recrystallization by rhenium segregation arising from neutron transmutation. Under fusion energy irradiation conditions, our framework predicts a substantial reduction of the effective recrystallization temperature relative to unirradiated microstructures, while Re production restores and even elevates this limit.
By providing quantitative projections of recrystallization kinetics and in-service recrystallization temperatures, this work establishes a predictive tool for assessing the lifetime and operational envelope of W-based plasma-facing materials under fusion conditions.
\end{abstract}



\section{Introduction}
Tungsten (W) has been identified as the leading candidate for plasma-facing components (PFCs) in fusion devices owing to its advantageous properties, including an exceptionally high melting point, excellent thermal conductivity, low tritium retention, small thermal expansion coefficient, low neutron activation, resistance to erosion, and high yield strength at elevated temperatures.  
Despite these benefits, W suffers from unacceptably high values of the ductile-to-brittle transition temperature (DBTT), which can range between room temperature for high-purity, cold-rolled specimens \cite{reiser2016ductilisation,bonnekoh2019brittle,brunner2000analysis} to over 500$^\circ$C for conventionally forged specimens \cite{Kim2014,butler2018mechanisms,ren2019investigation}.  
Due to the intermittent --but cyclic-- nature of the plasma kinetics during fusion reactor operation, it is also important to consider elements of thermal fatigue in the mechanical performance of W components \cite{mcelfresh2024fracture,pintsuk2013qualification,pintsuk2015characterization,loewenhoff2012tungsten,wang2016thermal,fukuda2020effect,shah2021numerical,ma2023high}. 

The DBTT of conventional polycrystalline tungsten depends on the material microstructure, and can be lowered by plastic deformation to values even below room temperature \cite{reiser2012tungsten,reiser2016ductilisation,pantleon2021thermal}. Pre-straining by, e.g., cold working or rolling is the basis for the processing of the so-called `fusion-grade' tungsten \cite{terentyev2017mechanical,yin2018tensile}. However, the deformation structures resulting from low-temperature deformation are thermally unstable, and predispose the material to a set of restoration processes during operation at high temperatures: recovery, recrystallization, and grain coarsening. The unavoidable occurrence of one or more of these processes during fusion device operation ultimately brings W back to its inherently brittle undeformed state, effectively setting practical limits on the material's lifetime. Recrystallization (RX) and grain growth are enabled by grain boundary (GB) motion, driven by differential driving forces resulting from chemical potential jumps across the boundary \cite{mcelfresh2023initial}, or from the anisotropy of the GB energy itself \cite{rohrer2023grain}. During fusion operation, discontinuities in dislocation and irradiation defect densities across grain boundaries, plus the surface tension ascribed to GB curvature, all create forces acting on grain boundaries and triple junctions. When GB mobility is sufficiently high, typically at elevated temperatures, these forces drive microstructural evolution leading to mechanical property changes. 
It stands to reason that a natural strategy to delay grain coarsening is then to promote Zener pinning by selective alloying and GB segregation \cite{lang2021recrystallization,gietl2022neutron} (notwithstanding other detrimental processes that might result from alloying). 
For example, the recrystallization temperature limit of K--doped W--3Re alloys was reported to increase beyond the 850$^\circ$C value measured for pure W and undoped W--Re \cite{gietl2022neutron}.

Neutron irradiation adds an extra dimension to microstructural evolution under fusion operation in tungsten PFCs.  
Irradiation induces both displacement damage and nuclear transmutation, leading to defect cluster accumulation and internal chemistry changes, often in a compounded manner \cite{Hasegawa2016,Tanno2007,fukuda2012effects,Hu2016,Garrison2017}. Irradiation-induced RX is a well-known phenomenon, particularly in tungsten and other refractory alloys where low dislocation mobility and poor grain boundary (GB) cohesion often leave deformed microstructures in a metastable state, making them highly sensitive to recovery processes \cite{vaidya1983radiation,mathaudhu2009microstructures,gietl2022neutron,Tyumentsev2025,Rest2005}.
However, while irradiation effects in W have been widely investigated by modeling and simulation \cite{marian2012modeling,huang2017mechanism,marian2017recent,becquart2010microstructural,MARIAN2012293,castin2017onset,ZHANG2023101443,jin2018breaking,castin2018object,mason2019atomistic,hou2021influence,ma2024initial,wu2025influence,mohamed2025investigation}, grain growth and recrystallization have been comparatively much less studied. Mannheim and co-workers pioneered a multiscale framework that couples a mean-field recrystallization model with a cluster dynamics approach~\cite{mannheim2018modelling,mannheim2019controlled,mannheim2019long,shah2021numerical}. However, their models are limited to microstructures defined by equiaxed grains with fixed GB misorientation and a constant nucleation rate. These conditions are often too simplistic to fully capture the coevolution of irradiation damage and grain growth kinetics. 

In this work, we develop a three-way coupling between plastic deformation, neutron irradiation, and grain growth/recrystallization using a mesoscale simulation approach that integrates finite element crystal plasticity (CPFE), stochastic cluster dynamics (SCD), and vertex dynamics (VD). CPFE is used to simulate hot working of W polycrystals, which loads grains with dislocations and predisposes the system for further microstructural changes downstream. SCD and VD are run concurrently, with the former simulating irradiation damage accumulation and VD evolving the polycrystal's microstructure according to the driving forces created during device operation.
Our model allows for full consideration of realistic microstructures with general grain size and misorientation distribution functions rather than being limited by simplifying assumptions typically made to study grain growth and recrystallization in polycrystals.

Ultimately, our approach is designed to make physics-based projections of the lifetime and upper temperature limit of W-based materials subjected to fusion conditions.
Note that the phenomenon studied here, i.e., premature recrystallization due to irradiation, is distinct from the common phenomenon of irradiation-induced grain coarsening. While the latter is triggered by grain boundary migration driven by curvature and irradiation-enhanced GB mobilities, irradiation-induced recrystallization involves the creation and growth of new, defect-free grains driven by stored energy from irradiation damage and other sources.

The paper is organized as follows. Section~\ref{sec:meth} outlines the theoretical framework and describes the numerical implementation in detail in Section \ref{sec:coupling}. In Section \ref{sec:cond} we discuss the deformation and irradiation conditions under which the simulations are carried out.
Section~\ref{sec:results} presents the simulation results, including microstructural characterization and grain growth behavior under different thermal, mechanical, and irradiation conditions. 
An in-depth discussion and the interpretation of our results are provided in Section~\ref{sec:disc}.  We conclude the paper with our most important conclusions in Section~\ref{sec:conc}.

\section{Theory and methods}\label{sec:meth}

The coupling between crystal plasticity and vertex dynamics has been demonstrated in past publications \cite{admal:po:marian:2017,admal:po:marian:2017b,mcelfresh2023initial}, while the integration of stochastic cluster dynamics with crystal plasticity was achieved in a series of papers with W and Fe as subjects of study \cite{yu2021coupling,yu2022physics,chatterjee2025spatially}. The coupling between SCD and VD is a new contribution of this paper, although some of its aspects have already been partially combined in a recent work by the authors~\cite{yu2025simulations} in the context of deuterium exposure of W armor surfaces. Next, we describe the theory behind each of the three simulations modules in detail, highlighting the connection points among them.

\subsection{Finite-element polycrystal plasticity}

\subsubsection{Single crystal kinematics}

For a deformable body occupying a volume $\Omega_0$ bounded by a surface $\partial\Omega_0$, 
a one-to-one mapping $\vec{x}\left(\vec{X}, t\right)$ is assumed to exist between the position of material points in their reference position $\vec{X}$ and their current position $\vec{x}$. The deformation gradient of this mapping, $\matr{F}=\partial\vec{x}/\partial\vec{X}$ is typically decomposed multiplicatively into plastic and elastic contributions, $\matr{F}^{\rm P}$ and $\matr{F}^{\rm E}$ as:
\begin{equation}\label{eq:md1}
\matr{F} = \matr{F}^{\rm E} \matr{F}^{\rm P}=\matr{I}+\nabla\otimes\vec{u}
\end{equation}
where $\vec{u}$ is the displacement vector. This decomposition is central to the finite kinematic theory of crystal plasticity \cite{admal:po:marian:2017,asaro1983crystal,gurtin2000plasticity,forest1998modeling}, which is built on the existence of an \emph{intermediate} configuration bridging the reference (i.e., undeformed) and the current (deformed) configurations. $\matr{F}^{\rm P}$ connects the reference and intermediate configurations through a stress-free transformation. $\matr{F}^{\rm E}$ then brings this intermediate configuration to its final state by deforming the underlying lattice. 

The rate of change of $\matr{F}$ can be written as:
\begin{equation}
\dot{\matr{F}}=\frac{\partial\dot{\vec{x}}}{\partial\vec{X}}=\frac{\partial\dot{\vec{x}}}{\partial\vec{x}}\frac{\partial\vec{x}}{\partial\vec{X}}=\matr{L}\matr{F}
\end{equation}
where $\matr{L}$ is the velocity gradient, defined as:
\begin{equation}
\matr{L}=\dot{\matr{F}}^{\rm E}{\matr{F}^{\rm E}}^{-1}+\matr{F}^{\rm E}\left(\dot{\matr{F}}^{\rm P}{\matr{F}^{\rm P}}^{-1}\right){\matr{F}^{\rm E}}^{-1}=\matr{L}^{\rm E}+\matr{F}^{\rm E}\matr{L}^{\rm P}{\matr{F}^{\rm E}}^{-1}
\label{capL}
\end{equation}
where $\matr{L}^{\rm E}$ and $\matr{L}^{\rm P}$ are the elastic and plastic velocity gradients, respectively. Note that, written in this fashion, $\matr{L}^{\rm P}$ is directly expressed in the intermediate configuration.

Dislocations glide on slip systems $\alpha = 1, 2, \ldots , N$, where each $\alpha$ defines a glide direction $\vec{s}^\alpha$ and a slip plane normal $\vec{n}^\alpha$. These vectors are also expressed in the intermediate configuration, and satisfy:
$$\|\vec{s}^\alpha\|=\|\vec{n}^\alpha\|=1;~~\vec{s}^\alpha\cdot\vec{n}^\alpha=0;~~ \vec{s}^\alpha,~\vec{n}^\alpha = {\rm constant}$$ 
From eq.\ \eqref{capL}:
\begin{equation}
\dot{\matr{F}}^{\rm P}=\matr{L}^{\rm P}\matr{F}^{\rm P} 
\end{equation}
For dislocation-mediated slip, the evolution of $\matr{F}^{\rm P}$ is governed by a set of shear rates $\dot\gamma^\alpha(\vec{X}, t)$ on individual slip systems via the flow
rule:
\begin{equation}
\matr{L}^{\rm P}\left(\vec{X},t\right):=\sum_{\alpha=1}^N\dot\gamma^\alpha\left(\vec{s}^\alpha \otimes \vec{m}^\alpha\right)
\label{lp}
\end{equation}
For consistency, a strain measure defined in the intermediate configuration is the elastic (Lagrange) strain:
\begin{equation}
\matr{E}^{\rm E}=\frac{1}{2}\left(\matr{C}^{\rm E}-\matr{I}\right) 
\end{equation}
where $\matr{C}^{\rm E}={\matr{F}^{\rm E}}^T\matr{F}^{\rm E}$ and $\matr{I}$ is the identity matrix. It can be shown that, when the Helmholtz free energy density of the solid is expressed as:
$$\Psi=\frac{1}{2}~\matr{E}^{\rm E}:{\cal C}:\matr{E}^{\rm E}$$
where ${\cal C}$ is a $4^{\rm th}$-order elasticity tensor (obtained with temperature dependence using eq.\ \eqref{eq:shear_modulus}), then a linear relation exists:
\begin{equation}
\matr{S}={\cal C}:\matr{E}^{\rm E}
\label{piola}
\end{equation}
where $\matr{S}$ is a symmetric stress tensor in the intermediate configuration (known as the $2^{\rm nd}$ Piola-Kirchhoff stress). $\matr{S}$ is related to the true stress $\matr{\sigma}$ as:
$$\matr{S}=J{\matr{F}^{\rm E}}^{-1}\matr{\sigma}{\matr{F}^{\rm E}}^{-T}$$
where $J=\det\matr{F}$.

\subsubsection{Material model}\label{sec:cp}

In our model, the shear rates follow a strain-rate sensitivity dependence on stress as:
\begin{equation}
\dot{\gamma ^\alpha} = \dot{\gamma}_0 \left(\dfrac{\tau^\alpha}{g^\alpha} \right) ^{\frac{1}{m}} 
\label{eq:dotgamma}
\end{equation}
where  $\dot{\gamma}_0$ is a reference slip rate, $m$ is the strain-rate sensitivity exponent, $g^\alpha$ is the slip resistance and $\tau^\alpha$ is the \emph{resolved} shear stress. It is generally convenient to work directly with the lattice vectors $\vec{s}^\alpha$ and $\vec{n}^\alpha$ (in the intermediate configuration), in which case $\tau^\alpha$ is obtained as \cite{gurtin2000plasticity}:
\begin{equation}\label{rss} 
\tau^\alpha=\vec{s}^\alpha\left(\matr{C}^{\rm E}\matr{S}\right)\cdot\vec{n}^\alpha
\end{equation}
$g^\alpha$ includes all the sources of resistance to dislocation motion in glide system $\alpha$:
\begin{equation}
g^\alpha= \tau^\alpha_{l}+\tau_f^\alpha+\tau_{G}^\alpha+\tau_{\rm irr}^\alpha
\end{equation}
where
\begin{itemize}
    \item[] $\tau^\alpha_{l}(T)$ is the lattice friction,
    \item[] $\tau_f^\alpha$ represents forest hardening,
    \item[] $\tau_{G}^\alpha$ accounts for grain size-limited dislocation glide,
    \item[] $\tau_{\rm irr}^\alpha$ is the slip resistance induced by the buildup of irradiation defects.
\end{itemize}
\begin{enumerate}
\item The temperature-dependent lattice friction stress is adapted from ref.~\cite{lim2015physically}:
\begin{equation}
    \tau^\alpha_{l}(T)
    = \tau_0
    \left[
        1 - \frac{kT}{2\Delta H_{\rm kp}}
        \ln \left( \frac{\dot{\gamma}_0}{\dot\gamma^\alpha} \right)
    \right]^2,
    \label{eq:tauf}
\end{equation}
where $\tau_0$ is the Peierls stress at 0~K, $k$ is Boltzmann’s constant, $\Delta H_{\rm kp}$ is the kink-pair formation enthalpy, and $\dot\gamma^\alpha$ is defined in eq.\ \eqref{eq:dotgamma}.

\item A Kocks-Mecking law is used to evolve the dislocation density in the material  \cite{mecking1981kinetics}: 
\begin{equation}
\dot{\rho}^\alpha = \left(k_1 \sqrt{\rho^\alpha} - k_2^\alpha \rho^\alpha\right)\sum_\alpha^N{\left\lvert \dot{\gamma}^\alpha \right\rvert}
\label{eq:kocks1}
\end{equation}
where $\rho^\alpha$ is the dislocation density in slip system $\alpha$, $k_1$ is a hardening parameter, $k_2^\alpha$ is a recovery parameter, and $\sum_\alpha^N{\left\lvert \dot{\gamma}^\alpha \right\rvert}$ is the aggregate total shear rate. 
$k_1$ and $k_2^{\alpha}$ are related according to the relation \cite{beyerlein2008dislocation}:
\begin{equation}
    \frac{k_2^{\alpha}}{k_1}
    = \frac{\chi b}{\Delta Q}
    \left[
        1 - \frac{kT}{c_1 b^3}
        \ln \left( \frac{\dot{\gamma}}{\dot{\gamma}_0} \right)
    \right],
    \label{eq:k2_relation}
\end{equation}
where $\chi$ is an interaction parameter, $\Delta Q$ is a normalized activation energy, and $c_1$ is a proportionality constant.
The evolving dislocation density is then used to calculate the forest hardening contribution:
\begin{equation}\label{eq8}
\tau_f^\alpha = \mu b\sqrt{\sum_\beta  h_{\alpha\beta}\rho^\beta}
\end{equation}
where $\mu$ is the shear modulus and $h_{\alpha\beta}$ is the hardening matrix (with its components listed in Table \ref{table:hdis}).

\item $\tau_{G}^\alpha$ is quantified by the shortest glide path across a given grain, $\lambda^\alpha$, as:
\begin{equation}
    \tau_{G}^\alpha=\frac{2\mu b}{\lambda^\alpha}
    \label{eq:tau_D}
\end{equation}
and
\begin{equation}
    \lambda^\alpha = \min_n\left\{\vec{s}^\alpha\cdot\vec{l}_n\right\}
    \label{lambda1}
\end{equation}
with $\vec{l}_n$ (in 2D, $n =1, 2$) are the principal axes of the \emph{gyration} tensor representing a given grain.
For a generic grain $i$ enclosing a set of $N$ spatial points $\{\vec{r}_i\}$ (typically, the mesh points contained within), the gyration tensor is defined as $\vec{R} = N^{-1}\sum^N_i \vec{r}_i\otimes\vec{r}_i$.
In this fashion, grains of arbitrary size are approximated as ellipsoids with an aspect ratio given by $l_1/l_2$ (where $l_1=\|\vec{l}_1\|$, $l_2=\|\vec{l}_2\|$), and the $\lambda^\alpha$ is taken as the maximum of the projection of the slip direction $\vec{s}^\alpha$ with the two principal axes. More details about this approach to obtain $\tau_{G}^\alpha$ can be found in ref.~\cite{mcelfresh2024fast}\footnote{One can also use shape descriptors ascribed to the gyration tensor to obtain useful information about each grain. For example, the grain size, $L_i$, can be obtained as the \emph{radius of gyration}:
$$L_i = \sqrt{\left(l_1\right)_i^2+\left(l_2\right)_i^2}$$
from which the average grain size in a polycrystal containing $M$ grains, $\bar{D}$, can be obtained as:
$$\bar{L}=\frac{1}{M}\sum_j L_j$$}.

\item Finally, the slip resistance resulting from the interactions between dislocations and irradiation defects follows a dispersed-barrier hardening model \cite{patra2012crystal,barton2013,cui2018coupled,chatterjee2025spatially}:
\begin{equation}\label{eq:tau_irr}
    \tau_{\mathrm{irr}} = \mu b \sqrt{\sum_j\eta_jN_jd_j},
\end{equation}
where $N_j$ and $d_j$ are the number density and size of defect cluster $j$, respectively. 
$\eta_j$ is the so-called hardening coefficient.
\end{enumerate}

\subsubsection{Polycrystal plasticity implementation}

The FECP model used here is adapted from a previously-developed diffuse-crystal interface plasticity model designed for polycrystalline bcc metals \cite{admal2017diffuse}. No modifications were made to the original FECP model but a brief summary is provided here for completeness. The diffuse-crystal interface model identifies polycrystals as regions of the configurational space with different levels of lattice rotation imposed by appropriate deformation gradients. The compatibility of the total deformation gradient, $\matr{F}_{\rm px}$, requires that grain boundaries be seen as formed by a subclass of geometrically necessary dislocations (GNDs) that remove the incompatibility of the plastic rotation field. Note that $\matr{F}_{\rm px}$ (denoted by the subindex `px') is different to the deformation gradient defined in the previous section and is used only in the context of the creation of polycrystalline structures. Mathematically:
\begin{equation}\label{eqDecomp}
\matr{F}_{\rm px}^{\rm E}\left(\vec{X},0\right)=\matr{R}^0\left(\vec{X}\right), \quad\quad \matr{F}_{\rm px}^{\rm P}\left(\vec{X},0\right)=\matr{R}^0\left(\vec{X}\right)^T
\end{equation}
where $\matr{R}^0$ represents a piecewise-constant rotation field in the polycrystal that represents the misorientation across grain boundaries. Numerical discontinuities in $\matr{F}^{\rm E}_{\rm px}$ and $\matr{F}^{\rm P}_{\rm px}$ are avoided by implementing a smoothed step function in to capture jumps in the rotation field. The GND tensor can be obtained in the standard fashion as $\matr{G}=\matr{F}^{\rm P}_{\rm px}{\rm Curl}\matr{F}^{\rm P}_{\rm px}$.

$\matr{L}^{\rm P}$ (in eq.\ \eqref{lp}) is evolved in a piecewise manner in each grain due to the different orientations of the slip systems. The model is solved using a finite element approach in systems containing large numbers of grains. Details about the FECP model are given in refs.~\cite{admal:po:marian:2017a,mcelfresh2023initial}. 

\subsection{Vertex dynamics model for grain boundary evolution}

Our model of grain boundary evolution is based on a two-dimensional (2D) vertex model of the polycrystal microstructure \cite{mcelfresh2023initial}. The model consists of \emph{physical} nodes, representing triple junctions (TJ), and \emph{virtual} nodes connected as a piecewise discretization of a grain boundary. 
While 3D models of recrystallization and grain growth exist (see reviews in refs.\ \cite{raabe2001mesoscale,hallberg2011approaches}), 2D models have proven adequate to simulate microstructural evolution in highly-textured or longitudinally-deformed metals in the transversal direction \cite{anderson1984computer,anderson1989computer,jensen2009time,ma2004computer}. Indeed, this is mainly the case of fusion-grade W, which is generally rolled or extruded to increase GB cohesion and improve ductility.
A summary of the features of the model is provided next.

\subsubsection{Virtual node evolution equations:}\label{sec:virtual}

The motion of a grain boundary is defined by a generalized displacement $\mathbf{u} = (u_1 , u_2 , u_3 )$ with component $u_1$ representing motion in the direction of the GB normal, and $u_2$ and $u_3$ representing in-plane motion. The GB velocity, $\vec{v}=\dot{\vec{u}}$, responds to a set of driving forces defined by the vector $\vec{f}$:
\begin{equation}
\vec{v}=\matr{M}\vec{f}=\begin{pmatrix}
M_{11} & M_{12} & M_{13} \\
M_{12} & M_{22} & M_{23} \\
M_{13} & M_{23} & M_{33} 
\end{pmatrix}
\begin{pmatrix}
\psi \\
 \sigma_{12} \\
  \sigma_{13}
\end{pmatrix},\label{pepei}
\end{equation}
where $\matr{M}$ is the mobility matrix. The first component of $\vec{f}$, $f_1=\psi$, includes all `chemical' sources of motion. The second and third components of $\vec{f}$ are $f_2 = \sigma_{12}$ and  $f_2 = \sigma_{13}$, which are shear stresses along the $x_2$ and $x_3$ directions in the GB plane. These stresses lead to so-called \emph{coupled shear boundary} motion through the corresponding mobility components $M_{12}$, $M_{23}$, and $M_{13}$. 

For two-dimensional domains, eq.~\eqref{pepei} takes the form
\begin{eqnarray*}
v_1 &=& M_{11}\psi + M_{12}\sigma_{12},\\
v_2 &=& M_{21}\psi + M_{22}\sigma_{12}.
\end{eqnarray*}
Under stress-free conditions, and with the GB unit normal vector $\vec{n}$ defined as $n_1=\cos\phi$ and $n_2=\sin\phi$ (with $\phi$ the polar angle of the GB normal in a 2D Cartesian coordinate system), the associated mobilities can then be written as:
\begin{eqnarray*}
M_{11}=M(\theta,T)n_1,\\
M_{21}=M(\theta,T)n_2.
\end{eqnarray*}
Where $M_{11}$ captures motion of a GB vertex along the GB normal, and $M_{21}$ represents its motion in the GB plane. $M(\theta,T)$ is defined in Sec.\ \ref{sec:mob}. Motion governed by $M_{11}$ responds to a mechanical driving force, while motion defined by $M_{21}$ reacts to `configurational' driving forces. 
In this work, we disregard the latter, which leads to the following simplified equation of motion:
\begin{equation}\label{GB_velocity_2D}
    v_1 = M_{11}\psi.
\end{equation}

The total chemical driving force, $\psi$, acting on the GB is given by
\begin{equation}
    \psi = -\frac{\gamma(\theta)}{R} + \mu b^2 \Delta \rho + \Delta E^{\text{irr}}
    \label{eq:psi}
\end{equation}
which captures driving forces associated with GB curvature (note that here we do not include the effect of GB stiffness \cite{rohrer2023grain}), and with differential elastic and damage energies associated with discontinuities in dislocation and irradiation defect densities, respectively, across the GB.
In eq.\ \eqref{eq:psi}, $\gamma(\theta)$ is the GB energy as a function of misorientation angle $\theta$, and $R$ the local radius of curvature. $\Delta \rho = \rho_1 - \rho_2$ denotes the dislocation density difference and $\Delta E^{\text{irr}} = E^{\text{irr}}_1 - E^{\text{irr}}_2$ is the differential damage energy due to irradiation defect accumulation.
This energy can be expressed as
\begin{equation}\label{defectEnergy}
    E^{\text{irr}} = \sum_l\sum_n N_n^\nu~e^\nu(n) 
\end{equation}
where superindex $\nu$ refers to the defect type (`V' for vacancies or `SIA' for self-interstitial atoms) and $n$ indicates the number of point defects in a given cluster. The concentration of such a cluster, $N_n$, is defined as in eq.\ \eqref{eq:tau_irr}, while its size $d(n)$ is related to $n$ as:
\begin{equation}\label{eq:def-sizes}
    d(n)=\left\{\begin{matrix}\left(\frac{4n\Omega_a}{\pi b}\right)^{1/2},~{\rm if}~\nu\equiv{\rm `SIA'}\\
\left(\frac{3n\Omega_a}{2\pi}\right)^{1/3},~{\rm if}~\nu\equiv{\rm `V'}
\end{matrix}\right.
\end{equation}
where $\Omega_a$ is the atomic volume. The energy functions are defined for simplicity as $e^l(n) = e_0^l n^{2/3}$ for $n>3$. 
The numerical parameter $e_0^l$, as well as $e^l(1)$, $e^l(2)$, and $e^l(3)$, are determined from atomistic calculations of defect cluster energies as function of cluster size~\cite{ma2020multiscale} and are all given in Table \ref{table:param}.  

The signs of the second and third terms in eq.~\eqref{eq:psi} indicate that the GB moves toward regions with higher energies, leading to grains with lower dislocation and irradiation defect densities.

\subsubsection{Grain boundary and triple junction mobilities}\label{sec:mob}

The components $M_{11}$ and $M_{12}$ introduced in Sec.\ \ref{sec:virtual} can be obtained from the general expression for the grain boundary mobility tensor as~\cite{chen2019grain,yu2025simulations}
\begin{equation}
\label{eq:mobility}
    M_{1j} = \frac{f_0 A}{kT}
    \left(b_da_j\right)
    \exp{\left(-\frac{Q_{\rm GB}}{kT}\right)},
\end{equation}
where $f_0$ is a kinetic prefactor, $Q_{\rm GB}$ is the activation energy, $A$ denotes the GB area, and $a_j=b_d$ if $j=1$ or $a_j=h_d$ if $j=2$. $b_d$ and $h_d$ are the Burgers vector and step height, respectively, of the leading disconnection defining the GB geometry.
For W, the relevant parameters have been obtained from molecular dynamics simulations and are summarized in Table~\ref{table:mobility_parameters}~\cite{chen2019grain,mathew2022interstitial}. 
It is important to note that the mobility of a GB in W can be strongly influenced by segregation of Re produced by transmutation. 
Saturated Re segregation has been shown to increase the GB activation energy $Q_{\rm GB}$ by approximately 0.57~eV~\cite{zhang2020segregation}, thereby significantly reducing GB mobility. 

Equation \eqref{eq:mobility} results in nearly zero GB mobility at temperatures below 900~K and for misorientations less than $75^{\circ}$. Within the range where mobility becomes appreciable, Re-segregated boundaries exhibit values roughly two orders of magnitude lower than those of pristine W, indicating that Re significantly delays GB motion and hinders grain growth. 
In our simulations, the GB mobility for Re-segregated cases is assumed to correspond to the saturated limit and is linearly interpolated between the pristine and Re-saturated states as Re accumulates during irradiation.

\subsubsection{Triple junction motion}\label{tj2}
The influence of triple junctions (TJs) on grain growth during annealing has been highlighted in both theoretical analyses and simulation studies \cite{gottstein2005triple,gottstein2002triple,gottstein2004validity}. 
When the motion of these junctions controls grain evolution, the growth behavior no longer follows the conventional Johnson-Mehl-Avrami-Kolmogorov (JMAK) framework~\cite{price1990use}, nor its Mullins-Neumann modification~\cite{gottstein2004validity}. Both experimental evidence and numerical simulations confirm that TJs possess finite mobility, as seen in the persistent differences between dynamic and equilibrium dihedral angles during migration \cite{gottstein2001grain,gottstein2002triple,gottstein2005triple,thomas2019disconnection}, but display a pronounced temperature dependence. At high temperature, it has been shown that TJ mobility is generally not the rate-limiting step during grain evolution \cite{czubayko1998influence}, which will be the general assumption considered in this work.
Triple junction motion can be described by a viscous relation:
\begin{equation}
\mathbf{v}_{\rm TJ} = {\cal M}_{\rm TJ} \sum_{k=1}^{3} \gamma_k \mathbf{t}_k ,
\label{tj1}
\end{equation}
where $k$ indexes the three grain boundaries meeting at the junction, and ${\cal M}_{\rm TJ}$ denotes the TJ mobility. Here, $\gamma_k$ is the specific energy of GB `$k$', and the corresponding driving forces act along the tangential directions $\mathbf{t}_k$.  
At equilibrium, $\mathbf{v}_{\rm TJ}=0$. If the GB energies are assumed to be misorientation-independent, eq.~\eqref{tj1} reduces to the classical dihedral-angle (Herring) condition. A direct consequence is that grains bounded by angles smaller than $90^\circ$ are unstable and will eventually vanish.

When the dimensionless parameter $\Lambda = L{\cal M}_{\rm TJ}/M_{\rm GB}$ is large ($L$ being the grain size; note that the mobilities of TJs and GBs differ by a length dimension), as is the case here, the dihedral-angle condition is effectively preserved and the TJ rapidly adjusts to GB motion. In this work, the dimensionaless parameter $\Lambda$ is set to be constant at a value of 100 in all simulations.


\subsubsection{Grain boundary energies}

\begin{figure}[H]
    \centering


\begin{tikzpicture}
    \pgfplotsset{
        set layers, 
        every axis/.append style={
            scale only axis,
            width=10cm,
            height=6cm,
            xmin=0, xmax=90, 
        }
    }

    \begin{axis}[
    name=base,
        axis y line*=left,
        axis x line*=bottom,
        xlabel={\Large $\theta$~[$^\circ$]},
        ylabel={\Large $\gamma$ $\left[{\rm J}\cdot{\rm m}^{-2}\right]$},
        ylabel style={blue},
        yticklabel style={blue},
        grid=major,
        grid style={dashed, gray!30},
        yticklabel style={
    /pgf/number format/fixed,
    /pgf/number format/zerofill,
    /pgf/number format/precision=1
  }
    ]
    
        \addplot[blue, line width=2pt, no marks] 
            table [col sep=comma, x index=0, y index=1] {fig/misorientation/GB_energy_interpolated.csv};
        \label{plot:energy}
    \end{axis}

    \begin{axis}[
        axis y line*=right,
        axis x line=none, 
        ylabel={\Large $\lambda$ [nm]},
        ylabel style={red},
        yticklabel style={red},
    ]
    \draw[black, line width=0.8pt]
  (base.north west) -- (base.north east);
        \addplot[only marks,
  mark=*,
  red,
  mark size=2.8pt,
  mark options={
    fill=red,
    draw=red,
    line width=1pt
  }] 
            table [col sep=comma, x=x, y index=1] {fig/misorientation/lambda.csv};
        \label{plot:lambda}

    \end{axis}
\end{tikzpicture}
        \caption{Misorientation dependence of the GB energy, $\gamma(\theta)$, and denuded zone width, $\lambda(\theta)$.}
        \label{fig:parameters}
\end{figure}
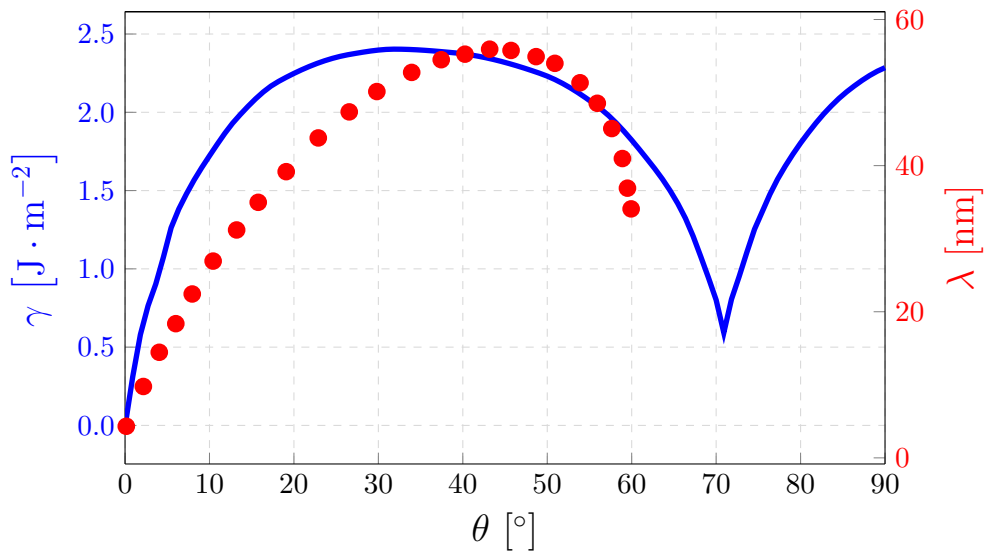

In this work, we take GB energies with a non-monotonic dependence on misorientation obtained using atomistic calculations. Specifically, we consider energies for $[110]$ symmetric tilt boundaries calculated using semi-empirical potentials \cite{frolov2018structures,feng2015energy,chirayutthanasak2022anisotropic} validated with electronic structure calculations \cite{setyawan2014ab,setyawan2012effects}. The $\gamma_{[110]}$-$\theta$ function for this family of GB is presented in Figure~\ref{fig:parameters}. 
The data displayed in the figure are then used to fit a set of Read-Shockley-Wolf functions (RSW) functions \cite{wolf1989read}:
\begin{equation}
    f_{\text{RSW}}(x,a)=\sin \left(\frac{\pi}{2x}\right)\left[1-a\log\left\{\sin\left(\frac{\pi}{2x}\right)\right\}\right]
    \label{eq:s-r-w},
\end{equation}
where $x=\left(\theta-\theta_{\text{min}}\right)/\left(\theta_{\text{max}}-\theta_{\text{min}}\right)$, and $\theta$ and $a$ are the misorientation angle and shape parameter \cite{sarochawikasit2021grain}, respectively. $\theta_{\text{min}}$ and $\theta_{\text{max}}$ represent the periodicity range of the $\gamma_{[110]}(\theta)$ function (0 to $90^\circ$ in this case), and are obtained by fitting eq.\ \eqref{eq:s-r-w} to atomistic calculations.

In our two-dimensional continuum grain growth model, the misorientation is calculated as the angular offset between the projected crystal axes of the grains on both sides of the GB. The inclination is then taken as the angle between the GB plane normal, $\vec{n}$, and these axes.

\subsection{Brief review of the stochastic cluster dynamics method}\label{sec:scd}

The accumulation of defects during irradiation is simulated using the stochastic cluster dynamics (SCD) model. SCD is a stochastic variant of the mean-field rate theory (MFRT) technique which relies on stochastic sampling of the underlying master equation for defect cluster evolution \cite{RN22,RN20}. Instead of deterministically solving exceedingly large sets of partial differential equations (PDE) of the concentrations of defects, as in standard MFRT, SCD evolves an integer-valued defect population $N_i$ in a finite material volume $\Omega$, thus avoiding exponential growth in the number of ODEs. This makes SCD ideal to treat problems where the dimensionality of the cluster size space is high, e.g., when multispecies simulations are of interest \cite{DUNN201643,MARIAN2012293,HE2026156222}.

SCD recasts the standard ODE system into stochastic equations of the form:
\begin{equation}\label{eq:scd}
\frac{dN_i}{dt}=\tilde{g}_i+\tilde{s}_{l}N_l- \left[\tilde{s}_{i}+\sum_\zeta\tilde{s}_{\zeta i}\right]N_i+
\sum_{j}N_j\left[\sum_{k}\tilde{K}_{jk}N_k - \tilde{K}_{ij}N_i\right]
\end{equation}
where the set $\{\tilde{g},\tilde{s},\tilde{k}\}$ represents the reaction rates for defect insertion, thermal dissociation, and annihilation at sinks, and binary reactions occurring inside $\Omega$. Subindices $j$ and $k$ refer to all distinct defect species present in the system, whereas the subindex $\zeta$ refers to the type of sinks available to absorb defects (`d' for dislocations, `gb' for grain boundaries, `ppt' for precipitates). The subindices in the expression satisfy $l-1=i$ and $j+k=i$ (reflecting dissociation by monomer emission and association of complementary clusters, respectively).
Equation \eqref{eq:scd} can be straightforwardly extended to partial differential equations to capture diffusion due to defect concentration gradients \cite{Yu_2020}.

Equation \eqref{eq:scd} is solved using the residence-time algorithm by sampling, selecting, and executing events from the set of rates $\{\tilde{g},\tilde{s},\tilde{K}\}$ \cite{RN22}. The volume $\Omega$ is in principle arbitrary, although, under certain conditions, it can be rescaled for improved computational efficiency. A quantitative demonstration of volume rescaling applied to a selected group of defect clusters is provided in ref.~\cite{HE2024155325}. For more information about the model, the reader is referred to past publications by the authors \cite{RN22,RN20,Yu_2020}. The time step for each SCD iteration is obtained {\it a posteriori} from a Poisson distribution with a rate parameter equal to $\left[\sum_{i,j}\left(\tilde{g}_i+\tilde{s}_i+\tilde{K}_{ij}\right)\right]^{-1}$.

Here we consider vacancies, self-interstitial atoms, and any cluster combination thereof. The microstructure of the irradiated W specimens will be discussed in Sec.\ \ref{sec:cond}, but it includes dislocations and grain boundaries acting as strong sinks for irradiation defects. The sink rates due to these two features, $s_{ij}$ in eq.~\eqref{eq:scd}, are written as $s_{ij} = s_\text{d} + s_\text{GB} = \left(k^2_{\rm d}+k^2_{\rm GB}\right)D_i$, where $k^2_{\rm d}=\rho$ and $k^2_{\rm GB}=6\sqrt{\rho}/L$ are the sink \emph{strengths}. Here, $\rho$ is the dislocation density (same as in Sec.\ \ref{sec:cp}), $L$ is the grain size (same as in Secs.\ \ref{sec:cp}, \ref{sec:virtual}, and \ref{sec:mob}), and $D_i$ is the diffusivity of defect species $i$. In polycrystalline tungsten, each grain $k$ is assigned its own dislocation density $\rho^{(k)}$ and grain size $L^{(k)}$, yielding a grain-specific sink strength:
\begin{equation}
    S_d^{(k)} = \left( \rho^{(k)} + \frac{6\sqrt{\rho^{(k)}}}{L^{(k)}} \right) D_i
    \label{eq:sink}
\end{equation}
Both $\rho^{(k)}$ and $L^{(k)}$ are time-dependent quantities that are updated as the simulations proceed.




\subsubsection{Effect of crystal orientation on irradiation damage accumulation}
To include a dependence on GB misorientation in the sink rate for a given grain boundary, we adopt a correlation derived from the analysis of void-denuded zones (VDZs) near grain boundaries in Cu irradiated at high temperatures \cite{han2012effect}. Han et al.\ measured a strong correlation between GB character and the width of the VDZs. A wide VDZ signals a high sink strength, while narrow or nonexistent VDZs indicate a low sink efficiency. On that basis, they extracted a correlation for the sink efficiency as a function of GB misorientation, $\lambda(\theta)$, shown in Fig.~\ref{fig:parameters}.
Here, for a given grain with size $L$ bounded by several different GBs with isorientations $\theta_k$, we apply an average value extracted as:
\begin{equation}
 \bar\lambda = \frac{\sum_k h_k \lambda_k(\theta)}{\sum_k h_k} 
 \label{eq:barlambda}
\end{equation}
where $k$ runs through all the grain boundaries surrounding the grain in question and $h_k$ are the lengths of each GB. We then replace the sink strength just defined for grain boundaries, $k^2_{\rm GB}$ with $k'^{2}_{\rm GB}=\bar\lambda k^2_{\rm GB}$. The parameterization and set of damage mechanisms included in SCD have been validated against dedicated ion-beam irradiation experiments at room temperature in W \cite{he2025parameter}.

\subsection{Three-way coupling among CPFE, SCD, and VD}\label{sec:coupling}

\begin{figure}[h!]
\resizebox{\textwidth}{!}{
    \boxed{
    \begin{tikzpicture}[
  font=\sffamily,
  >=Latex,
  node distance=10mm and 18mm,
  box/.style={
    rectangle, rounded corners=3pt,
    draw=black!60, very thick,
    align=center, text width=7.2cm,
    minimum height=12mm,
    fill=#1
  },
  sidebox/.style={
    rectangle, rounded corners=3pt,
    draw=black!60, very thick,
    align=center, text width=4.4cm,
    minimum height=12mm,
    fill=#1
  },
  decision/.style={
    diamond, aspect=2.2,
    draw=black!60, very thick,
    align=center, text width=3.6cm,
    inner sep=1pt,
    fill=#1
  },
  arr/.style={-Latex, very thick, draw=black!70},
  darr/.style={-Latex, very thick, draw=black!60, dashed},
  note/.style={align=center, font=\small, text=black!80}
]

\node (n1) [box=cyan!15] {Generate 2D polycrystal\\ from experimental\\ EBSD microstructures};

\node (n2) [box=cyan!15, below=of n1] {Construct vertex topology (triple junctions, GB nodes,\\ node connectivity)};

\node (n3) [box=white, below=of n2] {Volume-averaging of dislocation density field\\$\rightarrow$ obtain plastic strain energy density in each grain.};

\node (n4) [box=orange!18, below=of n3] {Update stresses in each grain; calculate GB curvature;\\ update irradiation defect concentrations};

\node (n5) [box=orange!18, below=of n4] {Update nodal positions and velocities};

\node (n6) [box=orange!18, below=of n5] {Remove small grains; merge nodes};

\node (d1) [decision=orange!18, below=14mm of n6] {Reached exit\\ condition?};

\node (n7) [box=gray!15, below=12mm of d1] {Exit simulation};

\node (cpfem) [sidebox=white, right=34mm of n3] {Run crystal plasticity\\ finite element module\\ (CPFEM)};

\node (scd) [sidebox=magenta!12, right=34mm of n4] {Run stochastic\\ cluster dynamics\\ (SCD)};

\draw[arr] (n1) -- (n2);
\draw[arr] (n2) -- (n3);
\draw[arr] (n3) -- (n4);
\draw[arr] (n4) -- (n5);
\draw[arr] (n5) -- (n6);
\draw[arr] (n6) -- (d1);
\draw[arr] (d1) -- node[right=2mm, font=\small]{Yes} (n7);

\coordinate (loopL)  at ($(n4.west)+(-18mm,0)$);   
\coordinate (loopTop) at ($(loopL |- d1.west)$);   
\coordinate (loopMid) at ($(loopL |- n4.west)$);   

\draw[arr]
  (d1.west)
  -- node[below left=1mm, font=\small]{No} (loopTop)
  -- (loopMid)
  -- (n4.west);

\draw[arr,dashed] (cpfem.west) -- ++(-10mm,0) |- (n3.east);
\node[note] at ($(cpfem.west)!0.55!(n3.east)+(0,8mm)$)
  {Spatial dislocation \\density field, $\rho(\vec{r})$};

\draw[arr,dashed] ($(n4.east)+(0,+5pt)$) -- ++(10mm,0) |- ($(scd.west)+(0,+5pt)$);
\node[note] at ($(n4.east)!0.55!(scd.west)+(0,9mm)$)
  {Grain size ($L$)\\dislocation density ($\rho$)\\sink efficiency ($\bar\lambda$)};

\draw[arr,dashed] ($(scd.west)+(0,-10pt)$) -- ++(-10mm,0) |- ($(n4.east)+(0,-10pt)$);
\node[note] at ($(n4.east)!0.55!(scd.west)+(0,-10mm)$)
  {Defect type ($N_i$)\\ defect size ($n_i$)};



\end{tikzpicture}}
}
\caption{Schematic flowchart illustrating the coupling vertex dynamics, stochastic cluster dynamics and crystal plasticity.}
    \label{fig:flowchartSCD_VD}
\end{figure}
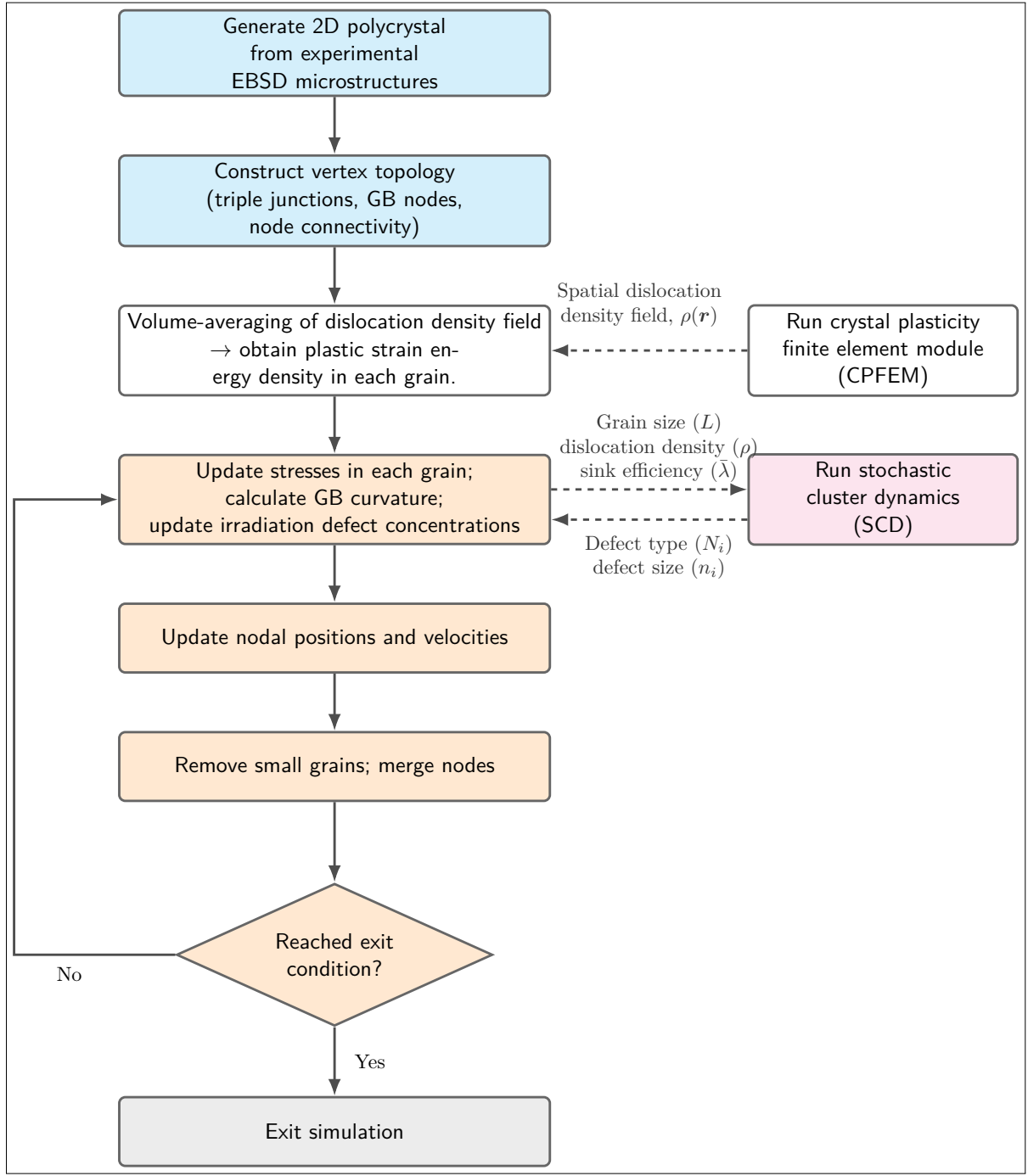
The computational framework integrates the crystal plasticity finite element simulations, stochastic cluster dynamics, and recrystallization simulations using the vertex dynamics model as illustrated in Fig.~\ref{fig:flowchartSCD_VD}. The vertex dynamics model acts as the overall microstructure simulator and dynamically passes information to (and receives it from) the other modules. The discretized initial vertex microstructure is used to calculate curvature-driven forces.
CPFE is run as a precursor step to simulate hot deformation and load the system with dislocations. The dislocation density field, $\rho(\vec{r})$, is then passed to vertex dynamics, which then homogenizes it in each grain by taking the spatial average:
$$\bar{\rho}_i=\Omega^{-1}\int_{\Omega}\rho(\vec{r})d\vec{r}$$
where $\Omega_i$ is the domain occupied by grain $i$. This set of $\bar{\rho}_i$ is then used to calculate the differential elastic energy driving forces. It is worth reemphasizing that the process simulated here, recrystallization, is governed by driving forces stemming from the differential accumulation of defects and dislocations on both sides of a GB, while standard grain growth is driven by overall reduction in GB area and GB energy. Thus, while the macroscopic structures at the end of a given simulation may look similar, the governing mechanisms are fundamentally different.

SCD is run concurrently with vertex dynamics, taking the values of $\bar{\rho}_i$ and $\bar{\lambda}_i$ as input to define the corresponding sink strengths and simulate neutron irradiation. SCD passes back defect concentrations and sizes, needed to calculate the differential damage energy driving forces on GBs.
As grain boundaries migrate, they sweep over dislocations and irradiation defects. At that point, the swept volume is considered to be defect-free, leading to a gradual reduction in the average dislocation density and defect concentration of the growing grain:
As the grain grows, the dislocation density is updated from iteration $(n)$ to iteration $(n+1)$ as:
\begin{equation}
    \bar\rho_i^{(n+1)} = \bar\rho_i^{(n)} \frac{\Omega_i^{(n)}}{\Omega_i^{(n+1)}}
    \label{eq:homogenize}
\end{equation}
and likewise for the defect concentrations.
Conversely, the dislocation density and defect concentrations remain unchanged in shrinking grains. This process is repeated until a suitable exit condition is achieved, e.g., the position of the nodes does not change from iteration to iteration within a given tolerance.
    
\section{Simulation conditions}\label{sec:cond}

\subsection{Crystal plasticity simulations}

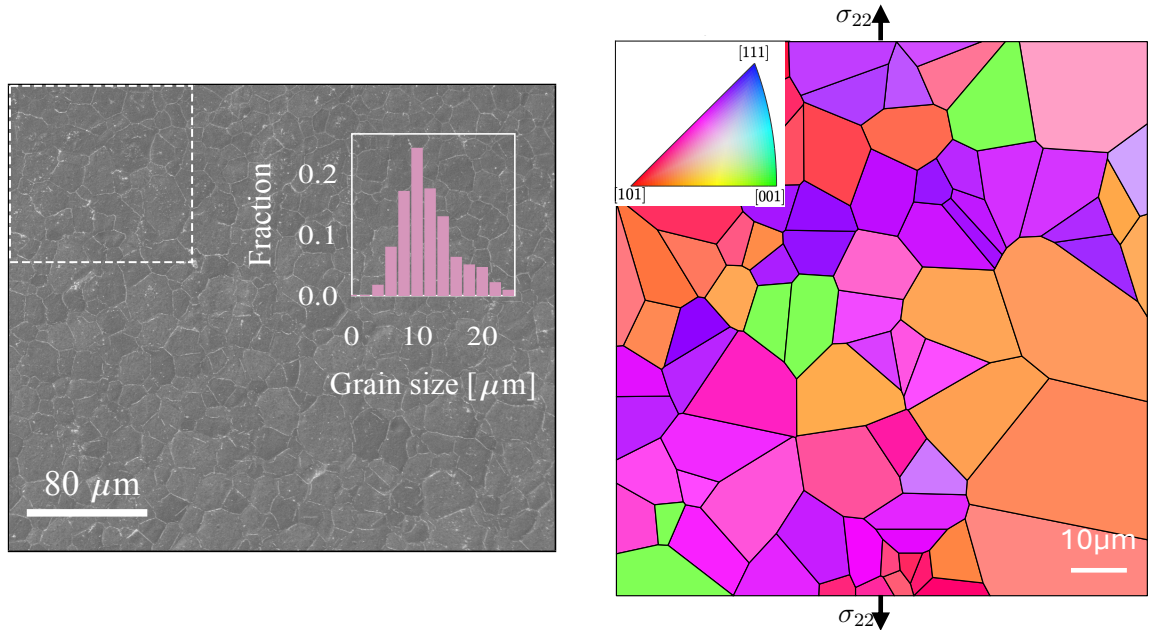
\begin{figure}[H]
\centering
 \raisebox{-0.5\height}{%
  \resizebox{0.50\textwidth}{!}{
  \includegraphics[width=0.99\linewidth]{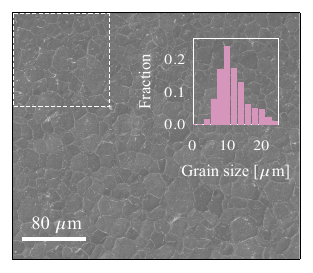}
  }
  }
\raisebox{-0.5\height}{%
  \resizebox{0.45\textwidth}{!}{%
    \begin{overpic}[width=0.5\linewidth]{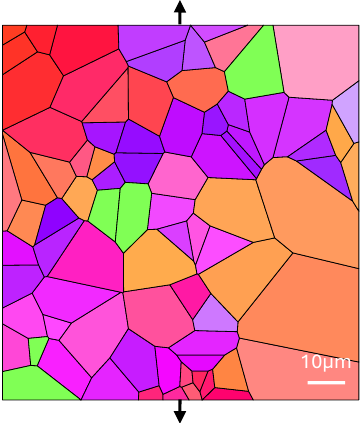}
      \put(0,180){\includegraphics[width=0.16\linewidth]{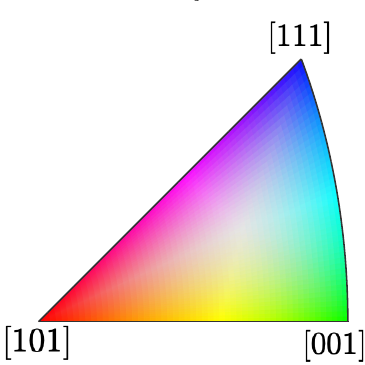}}
  \put(1,182){\tikz \draw[line width=0.35pt] (5.63cm, 2.4cm) -- (5.63cm, 0cm);}
  \put(1.4,250){\rule{73.5pt}{0.35pt}}
      \put(103,5){\color{black}\makebox(0,0){\large $\sigma_{22}$}}
      \put(102,260){\color{black}\makebox(0,0){\large $\sigma_{22}$}}
    \end{overpic}%
  }%
}%
\caption{Microstructural details of the polycrystal considered in this work. The left figure shows a cross-sectional view of the as-processed material obtained via scanning electron microscopy. Superimposed is the histogram of grain sizes. The dashed box indicates a cutout of a subspecimen similar to that used in the simulations. The right figure shows a reconstruction from EBSD analysis of the polycrystal employed in the simulations. The grain orientation of the material is colored according to the referential standard triangle shown in the inset. The arrows indicate the direction of loading in the CPFE simulations.}
\label{fig:CP-sims}
\end{figure}

\begin{figure}[ht]
\centering

\subfigure[GB misorientation map\label{fig:misorientation}]{
  \begin{minipage}[t]{0.45\textwidth}
    \centering
    \includegraphics[width=\linewidth]{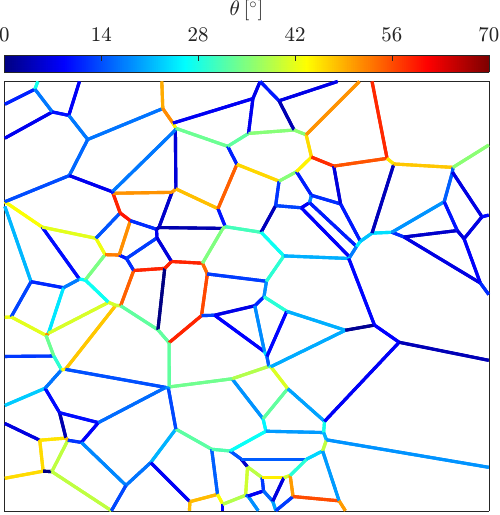}
  \end{minipage}
}
\hfill
\subfigure[Grain size distribution\label{fig:grainSizeDistribution}]{
  \begin{minipage}[t]{0.5\textwidth}
    \centering

\begin{tikzpicture}
\begin{axis}[
  width=7.5cm,
  height=7.5cm,
  xmin=0, xmax=40,
  ymin=0, ymax=0.16,
  xlabel={Grain size [$\mu$m]},
  ylabel={Fraction},
  tick align=inside,
  legend style={at={(0.98,0.98)}, anchor=north east, draw=black},
  legend cell align=left,
  yticklabel style={
    /pgf/number format/fixed,
    /pgf/number format/precision=2,
    /pgf/number format/fixed zerofill
  },
]

\addplot[
  ybar,
  bar width=0.9,
  fill=gray!45,
  draw=black,
  forget plot
] table[
  col sep=comma,
  x=BinCenter_um,
  y expr=\thisrow{HistogramPercent}/100
]{fig/Histograms/grain_size_all_data.csv};

\addplot[
  red,
  very thick
] table[
  col sep=comma,
  x=BinCenter_um,
  y expr=\thisrow{LogNormalFitPercent}/100
]{fig/Histograms/grain_size_all_data.csv};
\addlegendentry{Log-normal fit}

\end{axis}
\end{tikzpicture}
  \end{minipage}
}

\caption{Microstructural details of the polycrystalline specimen considered in the simulations.
\subref{fig:misorientation} Spatial grain boundary misorientation map.
\subref{fig:grainSizeDistribution} Grain size distribution.}
\end{figure}
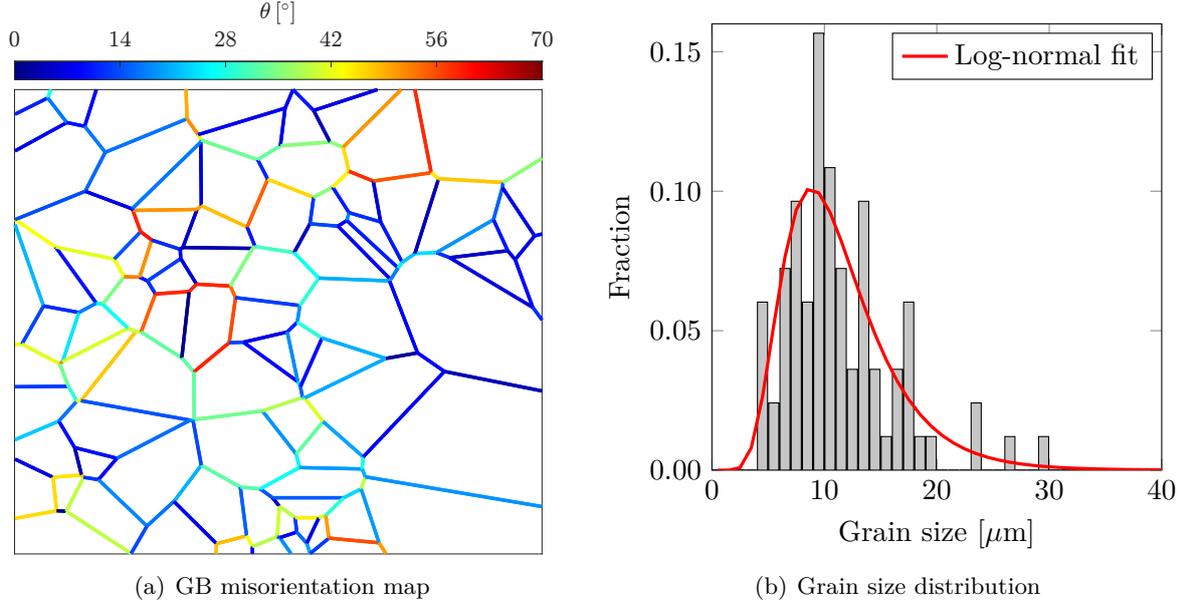

Synthetic two-dimensional microstructures were generated from electron back-scatter diffraction (EBSD) characterizations of pure tungsten specimens processed by sintering, hot-rolling, and annealing \cite{mcelfresh2024fracture,yu2025simulations}. The left image in Fig.~\ref{fig:CP-sims} shows a view of the surface of a specimen functionally equivalent to that considered here obtained with scanning electron microscopy. The figure on the right shows the EBSD map of a subsection of the original microstructure (represented by the dashed box in the left image), used here as a more tractable system from a computational perspective.
The EBSD map reveals a mild texture, with a modest predominance of grains oriented near the $[101]$ crystallographic direction. A subset of the full polycrystal was extracted to be used in the CP simulations, consisting of approximately 83 grains and 324 grain boundaries, as shown in Fig.~\ref{fig:CP-sims}. 
Figure~\ref{fig:misorientation} shows the corresponding GB misorientation function, consisting primarily of low-angle grain boundaries (LAGBs) ($\theta \le 15^\circ$) and a limited number of boundaries with $\theta \ge 45^\circ$. A log-normal fit was found to be an acceptable representation of the grain size distribution, as illustrated in Fig~\ref{fig:grainSizeDistribution}, with a peak at 10 $\mu$m and an average of approximately 11 $\mu$m, consistent with the size histogram given in Fig.\ \ref{fig:CP-sims}. 

The CP simulations are run under uniaxial deformation conditions maintaining grain compatibility at temperatures of 673 and 873 K and a strain rate of $10^{-3}$ s$^{-1}$. These values were chosen to replicate the experimental conditions found in ref.\ \cite{bonk2018cold}, which are used to calibrate the CPFE model and fit the value of $k_1$ (in eq.\ \eqref{eq:k2_relation}). The rest of the parameters are taken from ref.~\cite{terentyev2015dislocation} and are listed in Tables~\ref{table:hdis} and~\ref{table:param}. 
The numerical implementation uses 10,000 nodes with the Standard Lagrange (Quadratic) element type. The Crystal Plasticity Finite Element (CPFE) domain was discretized using a structured mapped mesh consisting of quadrilateral elements. The mesh resolution was determined by the grid sampling parameter, resulting in a regular Cartesian grid, where displacement field and dislocation densities were interpolated using quadratic Lagrange shape functions.
For the time-dependent study, the simulation is executed for a total time of $3 \times 10^1$ s with a temporal resolution of $\Delta t = 0.1$ s. 

\subsection{Irradiation conditions}\label{sec:irr}

\begin{figure}[H]
\centering
\subfigure[PKA energy distribution\label{fig:cpdf}]{

\begin{tikzpicture}
\begin{axis}[
  width=7cm,
  height=8cm,
  xmode=log,
  xmin=1e-3, xmax=1e3,
  ymin=0, ymax=1,
  xlabel={\Large $E_{\rm PKA}$ [keV]},
  ylabel={\Large $C(E_{\rm PKA})$},
  xtick pos=both,
  ytick pos=both,
  minor x tick num=9,
  grid=both,
  major grid style={draw=black!15},
  minor grid style={draw=black!15, dotted},
  ytick={0,0.2,0.4,0.6,0.8,1.0},
  scaled y ticks=false,
  yticklabel style={
    /pgf/number format/fixed,
    /pgf/number format/precision=1,
    /pgf/number format/fixed zerofill
  },
  legend style={
  draw=none,
  font=\small,
  at={(0.03,0.97)},
  anchor=north west
},
legend cell align=left
]

\addplot[green!80!gray, line width=2.8pt]
table[
  col sep=comma,
  x index=0,
  y index=1,
  skip first n=1
]{fig/irradiation/CPDFWDEMOhcll.csv};
\addlegendentry{DEMO-hcll}

\addplot[orange!80!gray, line width=2.8pt]
table[
  col sep=comma,
  x index=0,
  y index=1,
]{fig/irradiation/HFIR.csv};
\addlegendentry{HFIR}

\addplot[
  black,
  dashed,
  thick
] coordinates {
  (5.9,0)
  (5.9,1)
};

\node[
  anchor=west,
  font=\small
] at (axis cs:5.9,0.5) {$\bar{E}_{\rm PKA}$=5.9 keV};

\end{axis}

\end{tikzpicture}
}
\subfigure[Re transmutation rate\label{fig:ReConcentration}]{

\begin{tikzpicture}
\begin{axis}[
  width=7cm,
  height=8cm,
  ymin=0, ymax=2.5,
  xmin=0, xmax=10,
  xlabel={\Large dose [dpa]},
  ylabel={\Large Re amount [wt.~\%]},
  xtick pos=both,
  ytick pos=both,
  minor y tick num=9,
  grid=both,
  major grid style={draw=black!15},
  minor grid style={draw=black!15, dotted},
  xtick={0,2,4,6,8,10},
  scaled x ticks=false,
  xticklabel style={
    /pgf/number format/fixed,
    /pgf/number format/precision=1,
  },
  scaled y ticks=false,
  yticklabel style={
    /pgf/number format/fixed,
    /pgf/number format/precision=1,
    /pgf/number format/fixed zerofill
  },
  legend style={
  draw=none,
  font=\small,
  at={(0.03,0.97)},
  anchor=north west
},
legend cell align=left
]

\addplot[black!80!gray, line width=2.8pt] 
table[
  col sep=comma,
  x=dpa,
  y expr=\thisrow{Re_conc}*100
]{fig/irradiation/ReComposition.csv};

\end{axis}

\end{tikzpicture}
}
    \caption{\protect\subref{fig:cpdf} Cumulative PKA energy probability distribution function for neutron irradiation of W in the equatorial plane of the first wall armor structure in DEMO-hcll \cite{marian2025computational}. The average energy of the distribution is 5.9 keV. The distribution for HFIR irradiations (fission neutrons) \cite{ZHANG2023101443} is added for comparison \protect\subref{fig:ReConcentration} Transmutation rate of W into Re as a function of irradiation dose (from ref.\ \cite{gilbert2013neutron}). The graph stops at the accumulated dose after 1 year of continuous reactor operation.}
    \label{fig:irr_parameters}
\end{figure}
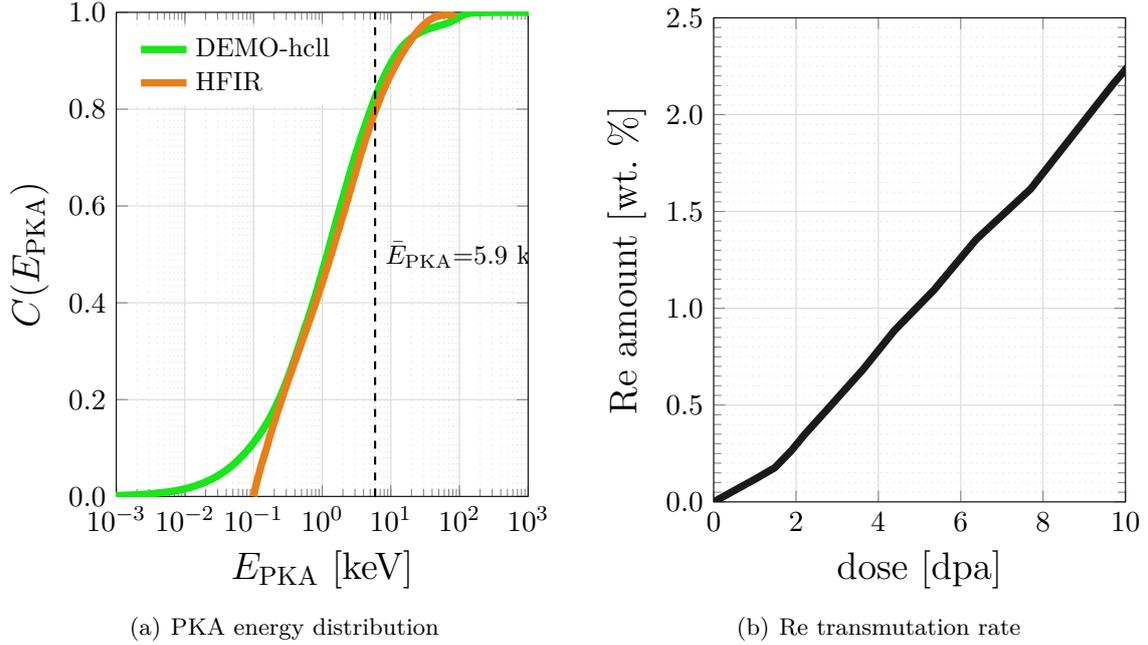

We simulate neutron irradiation of W polycrystals following the conditions expected in conventional fusion environments as defined by the He-cooled European DEMO concept with a liquid Pb-Li breeder (DEMO-hcll) \cite{gilbert2014comparative}. Key inputs for our simulations include:
\begin{itemize}
    \item The energy distribution of primary knock-on atoms (PKA), i.e., of lattice atoms that collide directly with incoming neutrons. This distribution is generally given in cumulative form, $C\left(E_{\rm PKA}\right)$. This distribution in turn depends on the neutron flux of the pertinent fusion reactor design, and is sampled iteratively to generate PKA with the correct energies. $C\left(E_{\rm PKA}\right)$ for the DEMO-hcll concept has been obtained by Marian \etal\ \cite{marian2025computational} and is given in Fig.~\ref{fig:cpdf}.
    \item The dose rate, typically given in \emph{displacements per atom} (`dpa') per second, given in Table \ref{tab:irr}. For W in DEMO-hcll, damage accumulates at a rate of 4.6 dpa per year \footnote{By contrast, no significant helium or hydrogen production is expected under these irradiation conditions.}.
    \item The temperature and the material microstructure.
    \item Nuclear transmutation through beta decay generates Re from W at a rate in DEMO-hcll provided in Fig.~ \ref{fig:ReConcentration}. The generated Re can segregate to the grain boundaries, leading to Zener pinning and slowing down GB motion.
\end{itemize}

\section{Results}\label{sec:results}

\subsection{Crystal plasticity simulations}\label{sec:cp-res}

\begin{figure}[H]
\centering
   \includegraphics[width=0.6\textwidth]{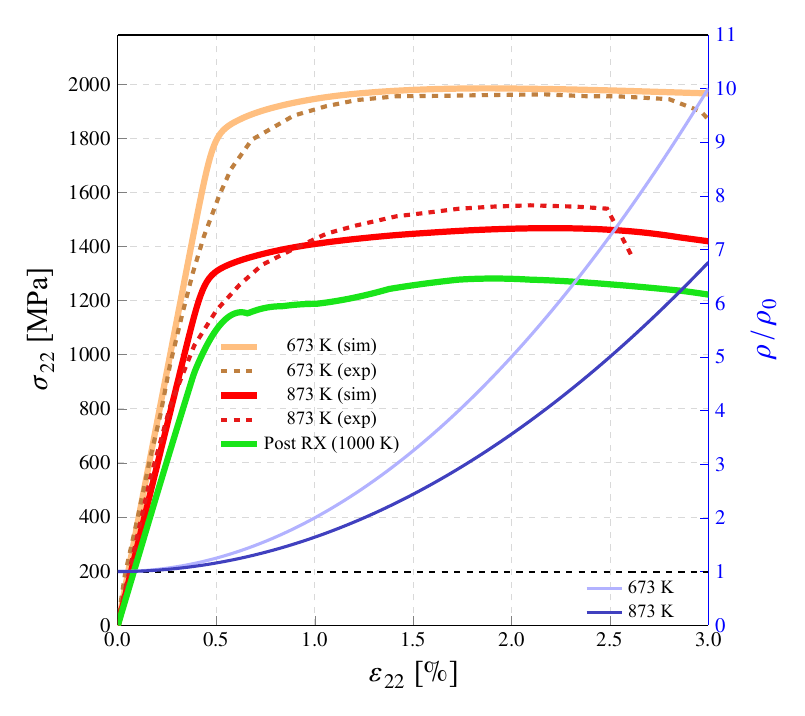}
\caption{Comparison of simulated and experimental~\cite{bonk2018cold} stress-strain curves at different temperatures (left $y$-axis) and associated dislocation densities (right $y$-axis). The stress-strain curve for the recrystallized specimen shown in Fig.\ \ref{fig:grainGrowth}(f) at 873 K is also shown. The buildup of dislocations is also shown at each temperature, normalized to the initial dislocation density of $\rho_0=10^{12}$ m$^{-2}$.} 
    \label{fig:stressStrain}
\end{figure}

The simulated stress-strain response of the specimen shown in Fig.\ \ref{fig:CP-sims} is given in Fig.~\ref{fig:stressStrain}. 
We use the experimental data by Bonk \etal~\cite{bonk2016cold} to fit the value of $k_1$ in eq.\ \eqref{eq:k2_relation}. The simulations are run up to 3\% total strain, which is the level at which some of the experimental specimens were seen to fail. As shown in the figure, we find that a value of $k_1=4.8\times10^6$ m$^{-1}$ provides excellent combined agreement at 673 and 873 K. The evolution of the dislocation density with strain is also shown in the figure in normalized units relative to the initial value of $\rho_0=10^{12}$ m$^{-2}$.

\begin{figure}[H]
\centering
\subfigure[Spatial dislocation density distribution\label{fig:spatial-density1}]{
  \begin{minipage}[t]{0.42\textwidth}
    \centering
  \raisebox{1.2ex}{  \includegraphics[width=0.97\linewidth]{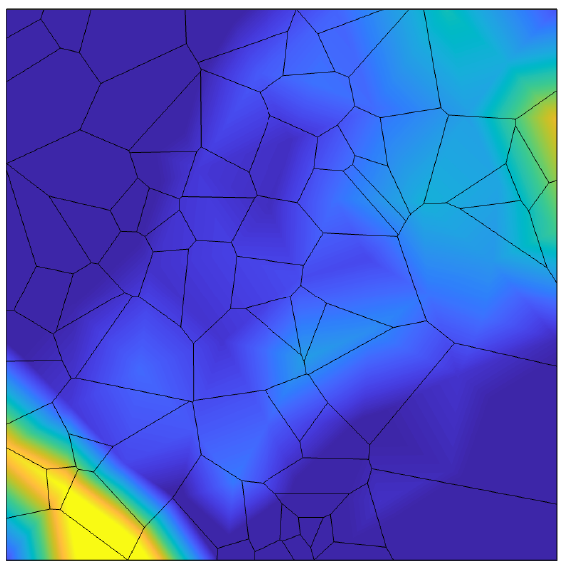} }
  \end{minipage}
}
\hfill
\subfigure[Homogenized dislocation density distribution\label{fig:spatial-density2}]{
  \begin{minipage}[t]{0.50\textwidth}
    \centering
    \includegraphics[width=\linewidth]{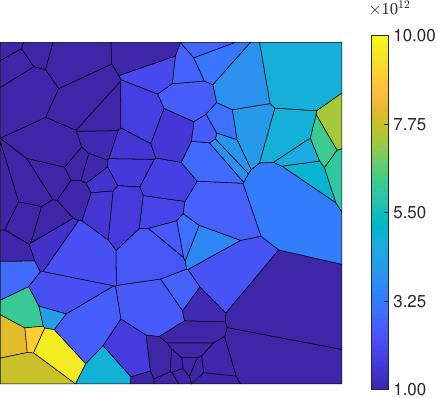}
  \end{minipage}
}
\caption{\protect\subref{fig:spatial-density1} Spatial dislocation density distribution after deformation to a strain of 3\% at 873 K.
\protect\subref{fig:spatial-density2} Corresponding grain volume-averaged dislocation densities. The color bar is common to both images and represents dislocation densities in units of m$^{-2}$.}
\end{figure}

The dislocation densities shown in Fig.\ \ref{fig:stressStrain} are not uniformly distributed throughout the polycrystal. Indeed, the spatial distribution of $\rho$ is highly heterogeneous, as shown in Fig.~\ref{fig:spatial-density1} for the polycrystal deformed at 873 K. The corresponding homogenized dislocation densities, obtained according to eq.\ \eqref{eq:homogenize}, are shown in Figure \ref{fig:spatial-density2}. These are the dislocation densities used to initialize the coupled SCD/RX simulations to be used to compute the differential elastic energy driving forces introduced in Sec.\ \ref{sec:meth}.

\subsection{Simulations of grain growth under neutron irradiation}

\begin{figure}[htb]
\begin{overpic}[width=1.0\linewidth]{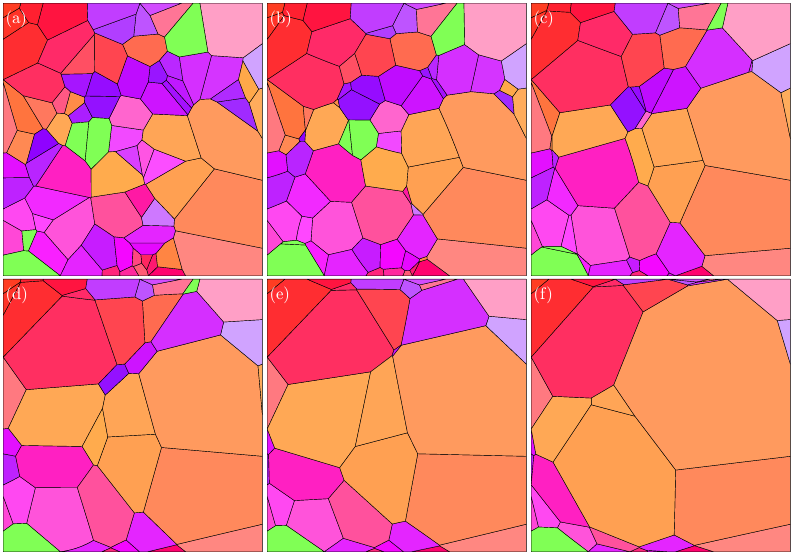}\hspace{-1.78em}
\end{overpic}  
\caption{\label{fig:grainGrowth}
Sequence of snapshots illustrating the evolution of the microstructure under irradiation at 1000 K. The color map follows that of Fig.\ \ref{fig:CP-sims}.}
\end{figure}

\begin{figure}[H]
\centering
\includegraphics[width=0.5\linewidth]{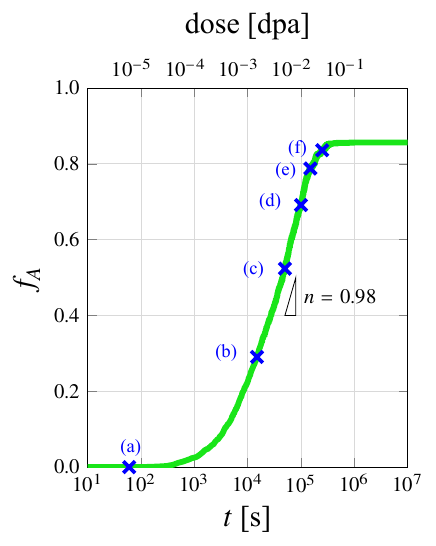}
\caption{Fraction of area transformed as a function of time. The '$\times$' markers represent the snapshots of the evolving microstructure given in Fig.\ \ref{fig:grainGrowth}. The time scale expressed as irradiation dose in dpa is shown on the top $x$-axis. The Avrami exponent obtained by fitting the numerical results to eq.~\ref{eq:avrami} is shown as the slope of the evolution curve.}
    \label{fig:f_A1}
\end{figure}

The microstructure shown in Fig.\ \ref{fig:spatial-density2} is subjected to irradiation under the conditions described in Sec.\ \ref{sec:irr} and allowed to evolve at a temperature of 1000 K. Figure \ref{fig:grainGrowth} shows a sequence of six snapshots of the evolving polycrystal (see the {\it Supplementary Movie} for an animated version of this sequence). Of special interest is the fraction of volume (or area) transformed as a function of time, indicating the completeness of the transformation. To quantify the transformed areal fraction, $f_A$, here we map the simulation domain onto a $100\times100$ uniform grid. 
A grid point is considered transformed once a GB sweeps over it at any point during the simulation. Each instance is counted only once, even if a given grid point is swept multiple times. 
$f_A$ is then defined as the ratio between the number of swept grid points and the total number of points in the domain. The time evolution of $f_A$ is provided in Fig~\ref{fig:f_A1}, with the location in the $f_A$-$t$ plot of the six snapshots shown in Fig.\ \ref{fig:grainGrowth}. 

As Figs.\ \ref{fig:grainGrowth} and \ref{fig:f_A1} show, the system takes 60 s to initiate its evolution (frame (a) in Fig.\ \ref{fig:grainGrowth}). 
This is akin to an \emph{incubation} time, i.e., a period where the irradiation defect density builds up to the level required to trigger GB and TJ motion. Subsequently, rapid growth takes place during the following 800 min of evolution (frames (b) to (c)), primarily driven by the coarsening of small grains consumed by larger ones throughout the domain. Beyond this stage, the growth rate gradually decreases (frames (d) to (e)), likely due to the emergence of a metastable or `locked' microstructure where only a few grains continue to grow. This transition manifests itself as a distinct inflection in the $f_A(t)$ curve at approximately $t$=$5.0\times10^4$~s. The system reaches saturation at a time of approximately $2.5\times10^5$~s ($\sim$69 hrs), frame (f), when nearly 85\% of the total area has been transformed.

To extract the dimensionality of grain growth, it is helpful to fit the evolution of $f_A$ to a modified JMAK expression \cite{fanfoni1998johnson,shirzad2023critical}, given in eq.\ \eqref{eq:avrami}, defined by the so-called Avrami exponent, $n$, which describes the dimensionality of the dominant kinetics.
For a grain growth process purely defined by GB motion, a value of $n\approx2$ is expected. For the plot in Fig.\ \ref{fig:f_A1}, we find $n=0.98$.

\subsubsection{Quantitative analysis of the recrystallized grain substructure}
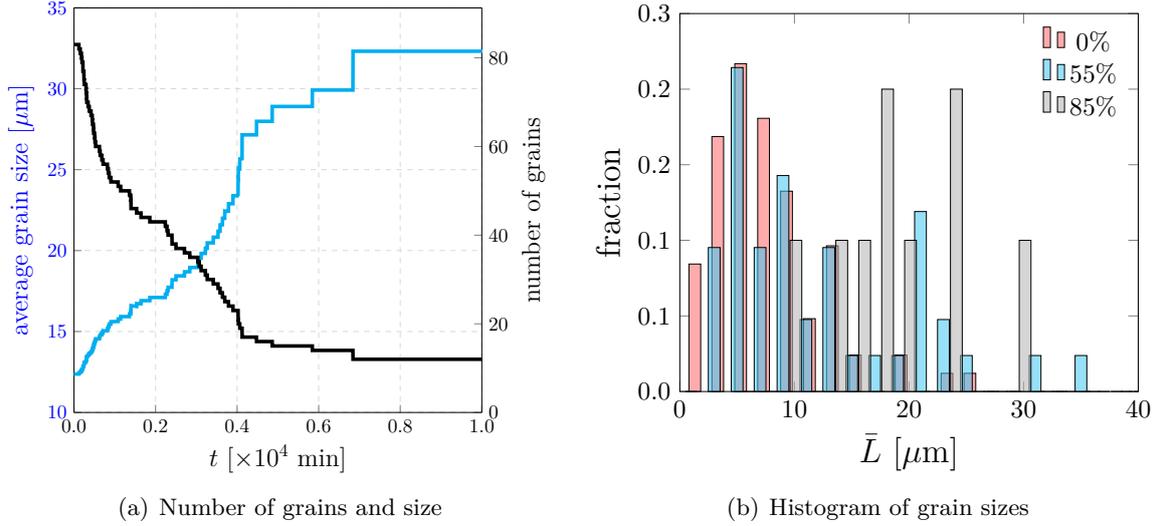
\begin{figure}[H]
\centering
\subfigure[Number of grains and size\label{fig:grainSize}]{
\resizebox{0.46\textwidth}{!}{


\begin{tikzpicture}
    \pgfplotsset{
        set layers, 
        every axis/.append style={
            scale only axis,
            width=8cm,
            height=8cm,
            xmin=0, xmax=1.0,
        }
    }

    \begin{axis}[
    name=base,
        axis y line*=left,
        axis x line*=bottom,
        xlabel={\Large $t$ [$\times10^4$ min]},
        ylabel={\Large average grain size~[$\mu$m]},
        ylabel style={blue},
        yticklabel style={blue},
        grid=major,
        ymin=10,ymax=35,
        grid style={dashed, gray!30},
        yticklabel style={
    /pgf/number format/fixed,
    /pgf/number format/zerofill,
    /pgf/number format/precision=0
  },
  xticklabel style={
    /pgf/number format/fixed,
    /pgf/number format/zerofill,
    /pgf/number format/precision=1
  }
    ]
    
        \addplot[cyan, line width=2pt, no marks] 
            table [col sep=comma, x expr =\thisrow{Time_s}/60/1e4, y=AvgGrainDiameter_um] {fig/Evolution/grainSizeEvolution.csv};
    \end{axis}

    \begin{axis}[
        axis y line*=right,
        axis x line=none, 
        ylabel={\Large number of grains},
        ylabel style={black},
        yticklabel style={black},
        ymin=0
    ]
    \draw[black, line width=0.8pt]
  (base.north west) -- (base.north east);
        \addplot[black, line width=2pt, no marks] 
            table [col sep=comma, x expr =\thisrow{Time_s}/60/1e4, y=GrainCount] {fig/Evolution/grainSizeEvolution.csv};

    \end{axis}
\end{tikzpicture}
}
}
\subfigure[Histogram of grain sizes\label{fig:grainSizeHistogram}]{
\resizebox{0.48\textwidth}{!}{
\begin{tikzpicture}
\begin{axis}[
 ybar,               
bar width=\BarWi,
  xmin=0, xmax=40,
  ymin=0.0, ymax=0.25,
  enlarge x limits=false,        
  enlarge y limits=false,
  xlabel={\Large $\bar L$ [$\mu$m]},
  ylabel={\Large fraction},
  enlarge y limits=false,
  yticklabel style={
  /pgf/number format/fixed,
  /pgf/number format/precision=1,
  /pgf/number format/fixed zerofill
},
  tick align=inside,
  legend style={at={(0.98,0.98)},anchor=north east,draw=none,fill=none},
]


\addplot+[
  ybar, bar shift=\ShiftA, mark=none,
  draw=black, fill=red!60, fill opacity=0.55
] table[
  x =GrainSize_um,
  y expr=\thisrow{Count_0pct}/83,
  col sep=comma
]{fig/Histograms/grainSizeHistogram.csv};
\addlegendentry{0\%}

\addplot[
  ybar, bar shift=\ShiftB, mark=none,
  draw=black, fill=cyan!60, fill opacity=0.55
] table[
  x =GrainSize_um,
  y expr=\thisrow{Count_55pct}/42,
  col sep=comma
]{fig/Histograms/grainSizeHistogram.csv};
\addlegendentry{55\%}

\addplot+[
  ybar, bar shift=\ShiftC, mark=none,
  draw=black, fill=gray!60, fill opacity=0.55
] table[
  x =GrainSize_um,
  y expr=\thisrow{Count_85pct}/10,
  col sep=comma
]{fig/Histograms/grainSizeHistogram.csv};
\addlegendentry{85\%}


\end{axis}
\end{tikzpicture}
}
}
    \caption{\protect\subref{fig:grainSize} Evolution of the total number of grains present, and their average size, in the simulation domain during irradiation/recrystallization simulations. \protect\subref{fig:grainSizeHistogram} The grain size distributions for three separate values of $f_{A}$ (snapshots (a), (c), and (f) in Figs.\ \ref{fig:grainGrowth} and \ref{fig:f_A1}) clearly show a progressive coarsening as the simulation advances.}
    \label{fig:grain_evol}
\end{figure}

The general trend observed in the simulations presented in the previous section is that microstructural coarsening proceeds primarily through the absorption of smaller grains by larger ones. In this section, we provide a quantitative characterization of the main microstructural descriptors and their evolution during grain growth.
Figure~\ref{fig:grain_evol} provides details about the temporal evolution of the polycrystal under irradiation. Figure \ref{fig:grainSize} shows how the average grain size changes during the course of the simulation, ranging from approximately 12 microns to more than 30 microns in the final recrystallized state. The total number of grains is shown also in the figure, starting with the original 83 and reducing to a total of only 11 at the end of the simulation.

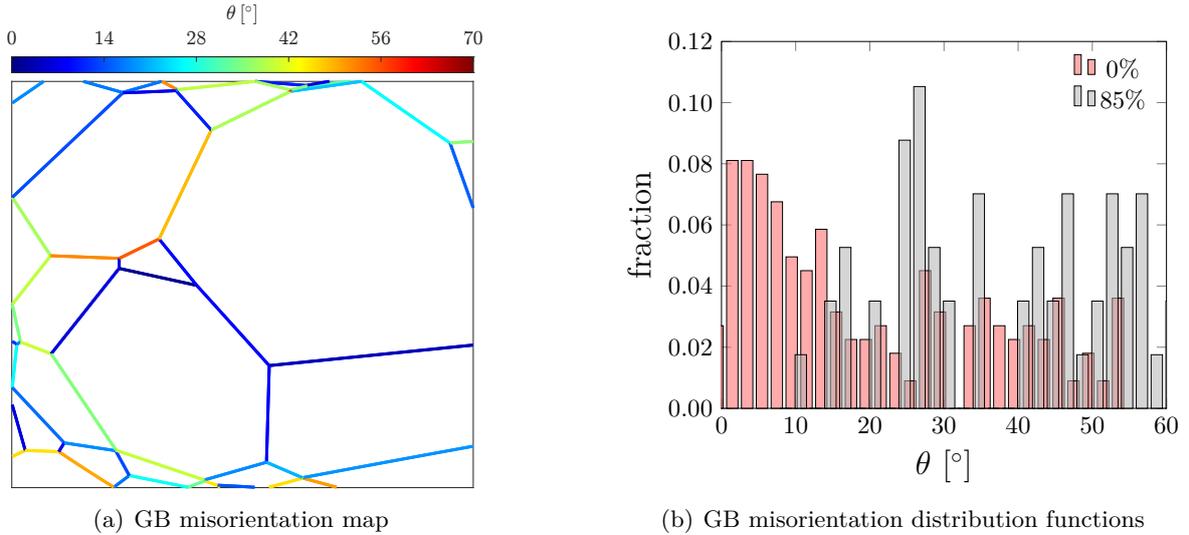
\begin{figure}[ht]
\centering
\subfigure[GB misorientation map\label{fig:misorientation85}]{
\resizebox{0.4\textwidth}{!}{
    \includegraphics[width=\linewidth]{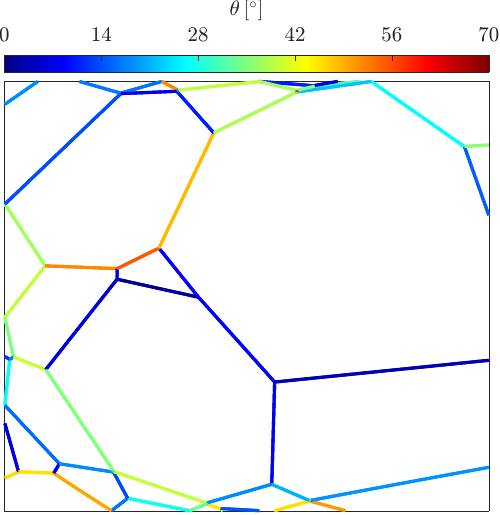}
}}
\hfill
\subfigure[GB misorientation distribution functions\label{fig:histo-miso}]{
\resizebox{0.49\textwidth}{!}{
\begin{tikzpicture}
\begin{axis}[
 ybar,               
bar width=\BarWi,
  xmin=0, xmax=60,
  ymin=0.0, ymax=0.12,
  enlarge x limits=false,        
  enlarge y limits=false,
  xlabel={\Large $\theta$ [$^\circ$]},
  ylabel={\Large fraction},
  enlarge y limits=false,
  yticklabel style={
  /pgf/number format/fixed,
  /pgf/number format/precision=2,
  /pgf/number format/fixed zerofill
},
  tick align=inside,
  legend style={at={(0.98,0.98)},anchor=north east,draw=none,fill=none},
]


\addplot+[
  ybar, bar shift=\ShiftA, mark=none,
  draw=black, fill=red!60, fill opacity=0.55
] table[
  x =Misorientation_deg,
  y expr=\thisrow{Fraction_0pct}/100,
  col sep=comma
]{fig/misorientation/misorientationHistogram.csv};
\addlegendentry{0\%}

\addplot+[
  ybar, bar shift=\ShiftC, mark=none,
  draw=black, fill=gray!60, fill opacity=0.55
] table[
  x =Misorientation_deg,
  y expr=\thisrow{Fraction_85pct}/100,
  col sep=comma
]{fig/misorientation/misorientationHistogram.csv};
\addlegendentry{85\%}


\end{axis}
\end{tikzpicture}
}}
\caption{Grain boundary misorientation distribution in the recrystallized state.
\subref{fig:misorientation85} Spatial grain boundary misorientation map.
\subref{fig:histo-miso} Grain misorientation distribution. \label{fig:misorientationAnalysis}}
\end{figure}

The grain size distributions at three representative stages of the process, corresponding to 0\%, 55\%, and 85\% completion (instances (a), (c), and (f) in Figs.\ \ref{fig:grainGrowth} and \ref{fig:f_A1}), are shown in Fig.~\ref{fig:grainSizeHistogram}. A clear coarsening trend can be detected as the system recrystallizes and small grains are consumed by the remaining larger grains. 
Likewise, the change in the misorientation distribution function between $f_A=0$ and $f_A=85\%$ is given in Fig.~\ref{fig:misorientationAnalysis}. Figure \ref{fig:misorientation85} shows the microstructure with the GBs highlighted by color in the same manner as the for initial microstructure in Fig.\ \ref{fig:misorientation}. Figure \ref{fig:histo-miso} gives the misorientation distribution functions of the recrystallized and original polycrystals. Note that the initial distribution corresponds to a highly-textured material (refer to the microstructure characteriation in ref.\ \cite{yu2025simulations}) and is thus not random. A clear shift from low- to high-angle GB prevalence is revealed, consistent with highly activated high-angle grain boundaries (HAGBs) sweeping through the material and with low-energy LAGBs being consumed in the process. These simulated misorientation distribution functions are in strikingly good agreement with those measured by Ren \etal~\cite{ren2019investigation}.

Finally, we carry out CPFE simulations of the final recrystallized microstructure using the previously calibrated Kocks-Mecking model at 873 K. In comparison with the initial microstructures, this one contains irradiation defects, a larger overall grain size, and a depleted total dislocation density. The interplay among these three features leads to the results added to Fig.~\ref{fig:stressStrain}, where a substantial decrease in flow stress is observed. Thus, our simulations predict the typical softening typically observed in hot-worked and annealed W specimens in the recrystallized state. 

\subsection{Sensitivity of microstructural evolution to key physical variables}

\begin{figure}[H]
\centering
\subfigure[Temperature ($\varepsilon$=3\%, $1.47\times10^{-7}$ dpa$\cdot$s$^{-1}$, no Re.)\label{fig:sens-temp}]{
    \includegraphics[width=0.45\linewidth]{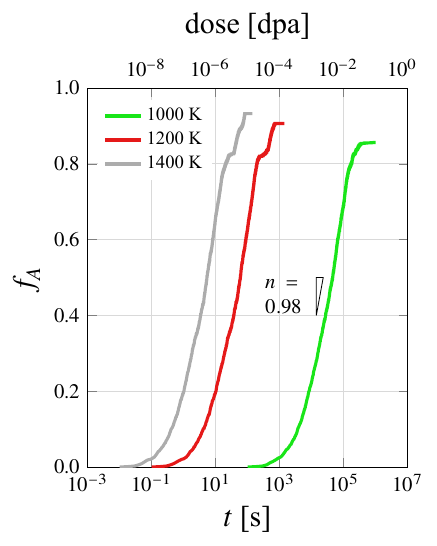}
}
\subfigure[Strain ($T$=1000 K, $1.47\times10^{-7}$ dpa$\cdot$s$^{-1}$, no Re.)\label{fig:sens-strain}]{
    \includegraphics[width=0.43\linewidth]{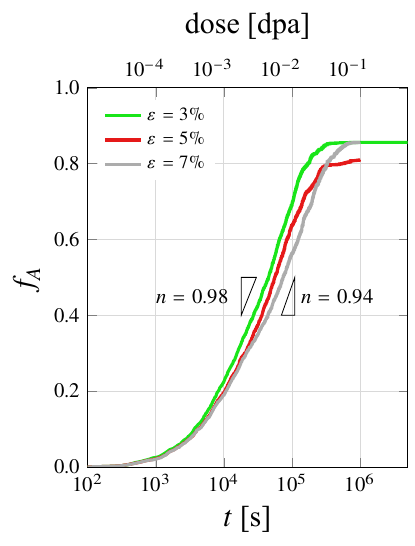}
}
\subfigure[Re content ($\varepsilon$=3\%, $T$=1000 K, $1.47\times10^{-7}$ dpa$\cdot$s$^{-1}$)\label{fig:sens-re}]{
    \includegraphics[width=0.45\linewidth]{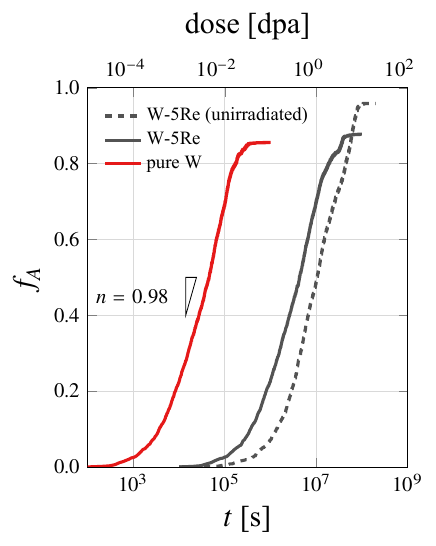}
}
\subfigure[Dose rate ($\varepsilon$=3\%, $T$=1000 K, no Re)\label{fig:sens-dose}]{
    \includegraphics[width=0.46\linewidth]{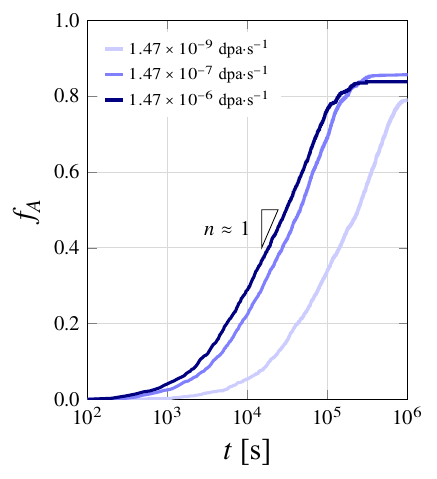}
}
    \caption{Sensitivity of the recrystallization process to key model variables. 
    (a) Temperature. (b) Total pre-irradiation strain. (c) Re content. (d) Dose rate. The values of the remaining parameters in each case are indicated in the subcaptions. Note that the time and dose scales are not the same in every plot.}
    \label{fig:sensitivity}
\end{figure}

Next, we study the kinetics of the recrystallization process to some of the key parameters of the simulations. Figure \ref{fig:sensitivity} shows the evolution of the transformed areal fraction as a function of temperature (Fig.\ \ref{fig:sens-temp}), pre-irradiation total strain (Fig.\ \ref{fig:sens-strain}), and alloy composition (Fig.\ \ref{fig:sens-re}). Starting with the effect of temperature in Fig.\ \ref{fig:sens-temp}, recrystallization and grain coarsening are drastically accelerated with increasing temperature, with incubation times of $10^5$~s (approximately 1.5 weeks) at 1000~K to approximately $100$~s and $10$~s at 1200 and 1400~K, respectively. This exponential sensitivity is primarily attributed to the thermally-activated character of GB mobilities, eq.\ \eqref{eq:mobility}, and should be regarded as one of the most important factors controlling RX in irradiated W. The Avrami exponent, however, is unaltered by changes in temperature, indicating no change in the underlying grain growth mechanisms, only in their rate.

The next variable to study is the total strain reached in the material after the deformation step conducted prior to irradiation. 
Pre-irradiation strain $\varepsilon$ is an important variable to study because it is ultimately linked to the total dislocation density accumulated after deformation. In Fig.\ \ref{fig:sens-strain} we study three cases, $\varepsilon$=3, 5, and 7\%. This is akin to the amount of cold working sustained by the material prior to irradiation. As the figure shows, strain mildly alters the kinetics of the transformation, with higher strain slightly delaying irradiation-induced grain growth. Interestingly, the Avrami exponent also decreases slightly, from 0.98 to 0.94, indicating some qualitative --not just quantitative-- differences in the transformation when the strain (or, equivalently, the dislocation density) increases.

The third variable under study is the alloy composition. Here, due to the lack of more detailed experimental information, we study the W-5Re system, which is the composition at which GB are expected to saturate with Re atoms \cite{zhang2020segregation,liu2020atomistic,gietl2022neutron}. As shown in Fig.~\ref{fig:sens-re}, Re pinning of GBs substantially delays the onset of RX compared to pure W. At 1000 K, pure W reaches 50\% recrystallized fraction after 45,000 s (12.6 hrs), compared to $4.3\times10^6$ s (1.6 months) in the case of W-5Re.
Thus, the presence of Re results in a delay of two orders of magnitude in time, extending the lifetime of the material from several hours to months. As another point of reference, unirradiated W-5Re reaches $f_A=0.5$ at $\sim$$10^7$ s (3.8 months). 
Our simulations thus confirm this beneficial aspect of neutron transmutation in W, notwithstanding other detrimental effects that the presence of Re may have on W-alloy performance.

Finally, we evaluated the effect of varying the dose rates on the transformation kinetics (all other irradiation parameters remaining the same). Figure \ref{fig:sens-dose} gives the dependence of $f_A$ on the dose rate at 1000 K. We have chosen a case where the dose rate is one hundredth of the nominal dose rate, which may be considered as representative of nuclear fission environments, and one that is 10 times higher than the nominal value, representative of fusion power plants.
As expected, higher dose rates accelerate the onset of RX due to the faster accumulation of irradiation defects, leading to stronger driving forces of the type described in Sec.\ \ref{sec:irr}.

\section{Discussion}\label{sec:disc}

\subsection{Parameter selection and physical consistency}

The present work involves models that require a total of 14 to 16 primary material parameters to characterize the various properties of tungsten. Needless to say, the fidelity of the coupled CPFE-SCD-VD framework relies on the selection of these parameters to bridge the atomistic scale with mesoscale evolution reliably. 
The majority of these parameters have been adequately vetted for use in models such as crystal plasticity \cite{cereceda2016unraveling} and stochastic cluster dynamics \cite{HE2026156222}, and here we focus on three parameters that have been relatively much less used in models of microstructure evolution.   

The first key parameter to discuss is the GB mobility, eq.\ \eqref{eq:mobility}. We have adopted a formulation containing a prefactor and an activation energy derived from molecular dynamics simulations \cite{mathew2022interstitial}. However, computing these mobilities is a daunting and challenging task, requiring substantial computational resources and able to only consider a limited set of simulation conditions. Here we employ the only set available for W computed to date, that of the [110] tilt boundaries. This by necessity constrains the range of possible evolutions in our simulations, which must therefore be analyzed in the context of this limitation. We do this in more detail in Sec.\ \ref{sec:avrami2} below.

The next physical parameter that merits discussion is the misorientation-dependent defect sink efficiency, $\lambda(\theta)$ defined in eq.\ \eqref{eq:barlambda}, which captures the heterogeneity associated with GB character of irradiation damage accumulation in the model. By utilizing a correlation based on void-denuded zone measurements, we move beyond the standard mean-field assumption of uniform sink strength and introduce a GB-induced variability that, while limited, reflects the microstructural complexity of irradiated polycrystals. This is particularly relevant for the transition from low to high-angle grain boundaries observed in Fig.\ \ref{fig:histo-miso}; the higher sink efficiency of high-angle boundaries accelerates the local removal of defects, thereby increasing the differential energy driving force that allows these grains to sweep through the deformed matrix. However, this sink strength variability is probably a minute subset of a more complex dependence that captures the multidimensional space of GB characters.

Lastly, the choice of the dimensionless parameter $\Lambda = 100$ for triple junction motion ensures that the system remains in the GB-migration-limited regime (see Sec.\ \ref{tj2}). While TJs can hinder growth at lower temperatures, our assumption that they do not act as the rate-limiting step at 1000 K is consistent with high-temperature vertex dynamics theory, where the Herring condition is effectively preserved. Nevertheless, the value of this assumption is largely unchecked, and more experimental and theoretical work is required to ascertain its relative importance.

\subsection{Phenomenology of recrystallization and implications for the model}

Unlike in most metals, deformed W can completely lose its ductility upon annealing at high temperatures. This is despite displaying a clear softening (generally measured in hardness) as temperature increases, in line with the typical behavior of most materials \cite{alfonso2014recrystallization}. It has been argued that textured W usually has a microstructure dominated by LAGB, whose high stability limits intergranular fracture pathways and confers the material a higher fracture toughness \cite{suslova2014recrystallization,ren2019investigation}. Upon high temperature deformation and recrystallization, however, HAGB are retained in the microstructure at the expense of LAGB. This facilitates intergranular crack propagation, partially explaining ductility loss in recrystallized W. This transfer from low to high-angle boundaries is consistent with our simulation results shown in Fig.\ \ref{fig:histo-miso}, as well as with the strength softening shown in Fig.\ \ref{fig:CP-sims} for the recrystallized microstructure.

The origin of LAGB in rolled W has been the subject of debate. Dislocations generated during deformation tend to form networks, and it has been reported that subgrains bounded by LAGBs may evolve from these networks \cite{farrell1967recrystallization,ekbom2002tungsten,reiser2020recrystallisation}. Some authors have proposed that the main factor in the recovery of hardness during RX is the disappearance and coalescence of low-angle grain boundaries formed by dislocation networks, followed by the formation of recrystallized nuclei at HAGBs and triple junctions \cite{farrell1967recrystallization,yu2016microstructure}. However, the presence of sub-boundaries in W created by dislocation arrangements is not conclusively established in literature studies.
Part of the challenge is that their identification is generally elusive, as most of the studies mentioned use EBSD to characterize the grain microstructure but subgrain boundaries do not lead to a distinctive signal in EBSD. This makes it difficult to separate them from the general GB substructure in our models, or to assign any special features to their behavior. Alfonso \etal~\cite{alfonso2014recrystallization} do not mention subgrain boundaries or invoke their presence to explain their results. Richou \etal.~\cite{richou2020recrystallization} simply presuppose their existence, designating sub-boundaries as any GB with a misorientation of less than 15$^\circ$. In addition, subgrains are typically associated with small grains formed within the original grains of the material. This would naturally lead to a separate peak in the grain size distribution function at small sizes \cite{zhang2016recrystallization}. The misorientation distribution function computed here in Fig.\ \ref{fig:histo-miso} is indeed bimodal, but there is no conclusive evidence that the low angle GBs can be unequivocally ascribed to smaller grains. 

Thus, while our CPFE model does not directly capture the formation of subgrains during the pre-straining stage prior to irradiation at high temperature, we believe that the impact of this limitation on microstructural evolution is modest. A stringer limitation might be that associated with considering only [110] tilt boundaries, which was the only orientation for which detailed atomistic data was available to us. As more information on general tilt (and twist) orientations becomes available, the model can easily incorporate it to increase its fidelity and applicability.

\subsection{Avrami exponents and kinetics of transformation}\label{sec:avrami2}

The most general form of the JMAK equation includes information about the incubation time, $t_0$, and the time to 50\% transformation, $t_{50}$, as:
\begin{equation}
f_A(t,T) =  1 - \exp\left[a\left(\frac{t-t_0}{t_{50}(T)}\right)^n\right]
\label{eq:avrami}
\end{equation}
where $a=-1$ is a constant \cite{shah2021numerical}. 
The Avrami exponent, $n$, is critical for characterizing the nature of a phase transformation. Specifically, $n$ provides information about the dimensionality and nucleation mechanism of the transformation process. 
In all the cases studied here (see Figs.\ \ref{fig:f_A1}, \ref{fig:sensitivity} and \ref{fig:sens-dose}), $n$$\approx$1, which is the value most frequently quoted in experimental studies of similar kind \cite{lin2012effects,alfonso2014recrystallization,wang2017effects,richou2020recrystallization} (a value of $n=2$ has also been reported \cite{ciucani2019recovery}). This adds confidence to our model predictions, particularly as it relates to the 2D vs.~3D difference between our simulations and actual experiments, which thus appears not to be a critical omission in RX of textured tungsten. Avrami exponents near unity are typically associated with (i) growth along a direction strictly perpendicular to grain boundaries, or (ii) when the rate of transformation is directly proportional to the amount of untransformed material remaining. Both of these are operative during microstructural evolution of the irradiated samples in this work.

\begin{figure}[H]
\centering
\subfigure[\label{fig:NoIrra}]{
\includegraphics[width=0.47\linewidth]{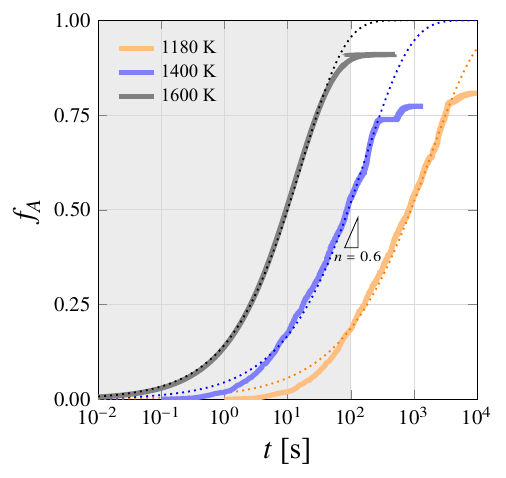}
}
\subfigure[\label{fig:t50}]{
\includegraphics[width=0.38\linewidth]{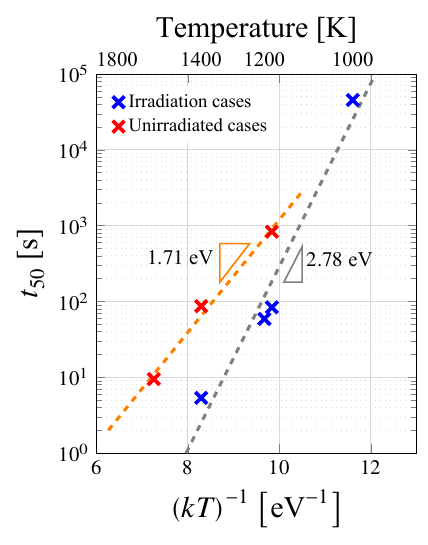}
}
\caption{\subref{fig:NoIrra} Areal transformation curves for 3\%-deformed W specimens as a function of temperature in the unirradiated condition. Taking the RX temperature as that at which $f_A=50\%$ at 100 s yields a value of 1400 K. Dashed lines illustrate the fits to the data using eq.\ \eqref{eq:avrami} with $n\approx0.6$. \subref{fig:t50} Semilogarithmic plot of the time to 50\% areal transformation as a function of temperature for W specimens deformed to 3\% total strain without and with irradiation. Fits according to eq.\ \eqref{eq:t50} are shown for both cases, with the respective activation energies shown for reference.}
    \label{fig:t50}
\end{figure}

We have also conducted tests without irradiation to evaluate the isolated effect of temperature on recrystallization. The results are shown in Fig.~\ref{fig:NoIrra} for 1180, 1400, and 1600 K. In this case, the Avrami exponent is significantly lower, $n\approx0.6$, in agreement with experimental data for cases where recovery and recrystallization overlap \cite{yu2017hardness,wang2017effects}. 
After Shah \etal~\cite{shah2021numerical}, the quantity $t_{50}$ in eq.\ \eqref{eq:avrami} is defined as the time to achieve 50\% microstructure transformation:
\begin{equation}
    t_{50}(T)=t_{50}^0\exp\left(\frac{\Delta E_{\rm RX}}{kT}\right)
    \label{eq:t50}
\end{equation}
The parameters $t_{50}^0$ and $\Delta E_{\rm RX}$ are obtained by fitting the data points extracted from Figs.\ \ref{fig:NoIrra} and \ref{fig:sens-temp} (for unirradiated and irradiated cases, respectively) as a function of temperature to expression \eqref{eq:t50}. The results of the fit are shown in Fig.~\ref{fig:t50} with the values of $\Delta E_{\rm RX}$ shown for reference. The complete fitted expressions for times in seconds are:
$$\begin{matrix}t_{50}^{\rm unirr}(T)=6.8\times10^{-5}\exp\left(\frac{1.71}{kT}\right)\\
t_{50}^{\rm irr}(T)=2.5\times10^{-10}\exp\left(\frac{2.78}{kT}\right)\end{matrix}$$
The values of $\Delta E_{\rm RX}$=1.71 and 2.78 eV obtained from the fits in eq.\ \eqref{eq:t50} represent effective activation energies of the recrystallization process. Several studies link the recrystallization temperature to the onset of self-diffusion on the basis that the activation energies extracted from the data are near the activation energy for vacancy migration in W ($\approx5.0$ eV) \cite{alfonso2014recrystallization,pantleon2021thermal,gietl2022neutron}. In irradiated alloys, the drop in $T_{R}$ is then attributed to radiation enhanced diffusion facilitating the motion of vacancies.
However, the measurements published in the literature (e.g., 8.3 eV \cite{denissen2006recrystallisation}, 7.5 eV \cite{mcelfresh2024fracture}, 5.9 eV \cite{alfonso2014recrystallization}, 4.5 eV \cite{Karanja2021Recovery}, 3.8$\sim$5.4 eV \cite{ciucani2019recovery}, 3.4 eV \cite{richou2020recrystallization}, and 3.7 eV \cite{reiser2020recrystallisation}) dispute this interpretation. Instead, we posit that recrystallization 
is mainly controlled by GB mobility. Atomistic calculations of GB mobilities in W do indeed result in GB activation energies ranging from 1.6 to 12.5 eV \cite{mathew2022interstitial}. As shown in Table \ref{table:mobility_parameters}, HAGBs with misorientations $>$80$^\circ$ display GB migration energies between 1.6 and 2.0 eV, consistent with the value of 1.71 eV obtained for the unirradiated cases in Fig.\ \ref{fig:t50}. In other words, HAGB are rate-limiting during deformation-driven recrystallization. 
Under irradiation, the larger value of $\Delta E_{\rm RX}$ ($\approx$2.8 eV) is likely a consequence of the shift to an evolution dominated by GB with misorientations around 70$^\circ$ (Table \ref{table:mobility_parameters}), either due to larger irradiation-induced driving forces on those GB, or to suppressed GB motion in those with $\theta>80^\circ$. 

For those reasons, we propose that the larger activation energies of between 5.5 and 6.0 eV measured experimentally is a reflection of GBs with thermally activated mobilities defined by those energies. Indeed, Table \ref{table:mobility_parameters} demonstrates the presence of GBs with energies around that range. Of course, here we have only considered mobilities with GBs with the [110] tilt axis, but it is reasonable to expect that similar activation energies would be found for other tilt orientations.

\subsection{Predictions of the recrystallization temperature}

Recrystallization is not an equilibrium phase transformation and thus the recrystallization temperature, $T_R$, is not a material constant but an operational quantity tied to a kinetic criterion. Generally, $T_R$ is defined as the temperature at which a prescribed fraction (commonly $f_A=0.5$) is reached after a fixed annealing time (e.g.\, $t=1.0$ hrs), or variants thereof \cite{WANG2024114162,alfonso2014recrystallization,Karanja2021Recovery}. $T_R$ can also be measured as the temperature at which the hardness drops by 50\% of the difference between the as-deformed and fully-annealed states \cite{alfonso2014recrystallization,richou2020recrystallization,shah2021numerical}. Note that the strength of the as-deformed material drops by 15\% in the recrystallized state in our simulations (see Fig.\ \ref{fig:CP-sims}). 

Here we choose to define it as the temperature at which the transformed areal fraction reaches 50\% after 100 s. With this in mind, here we conduct a series of simulations where we subject different microstructures to different conditions and systematically vary the temperature until finding the value for which $f_A=0.5$ after 100 s.
Table \ref{tab:T_R} shows the values of $T_R$ for specimens deformed to 3\% strain under three different irradiation conditions. Unirradiated specimens reach 50\% transformed fractions in 100 s at a temperature of 1400 K. Under irradiation, those specimens recrystallize at 1180 K. Accounting for Re production pushes the recrystallization temperature to 1550 K. Thus, our results predict the experimentally observed narrowing of the temperature window in its upper limit due to irradiation.
\begin{table}[H]
    \centering
        \caption{Recrystallization temperature simulated here for specimens deformed to 3\% strain subjected to different irradiation conditions. These values were obtained as the temperatures at which 50\% of the total specimen area was transformed after 100 s.}
    \begin{tabular}{|c|c|c|c|}
    \hline
      $\varepsilon$   & unirradiated & irradiated & irradiated (with Re) \\
    \hline
      3\%   & 1400 K & 1180 K & 1550 K \\
\hline
    \end{tabular}
    \label{tab:T_R}
\end{table}

$T_R$ values between 1350 and 1500 K have been quoted for unirradiated W polycrystals  \cite{gietl2022neutron,tsuchida2018recrystallization,zhang2016recrystallization,nikolic2018effect}, dropping to $\approx$1000 K after fission neutron irradiation to 0.42 dpa \cite{gietl2022neutron,ciucani2019recovery}. 
For its part, delayed RX due to the presence of Re in W is a well-known phenomenon \cite{golovanenko1976recrystallization,ramalingam1986elevated,tsuchida2018recrystallization,gietl2022neutron,shi2024effects}. Recrystallization temperatures of 1588$^\circ$C have been reported for W-0.5\%(wt.)Re-K \cite{shi2024effects}, 1500$^\circ$C for W-1\%(wt.)Re \cite{golovanenko1976recrystallization}, or 1650 to 1800 K in W-5Re (at\%) \cite{ramalingam1986elevated}. For irradiated W-5\%(wt.)Re, Gietl \etal~\cite{gietl2022neutron} found that a value 1473 K would adequately represent the recrystallization temperature, in very good agreement with our predictions.

\subsection{Alloy design}

\begin{figure}[H]
\centering
    \fbox{\includegraphics[width=0.55\linewidth]{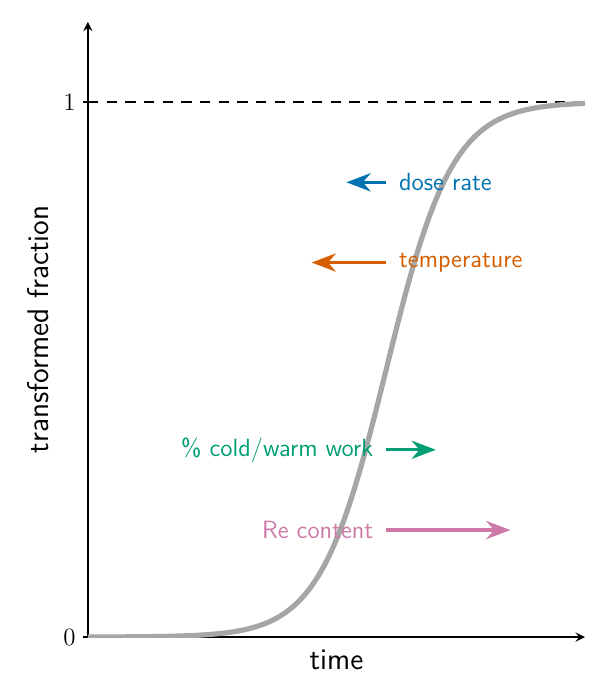}}
\caption{Schematic diagram showing the direction in time of different operational variables. The direction of the arrows indicates whether a given variable delays (pointing right) or accelerates (pointing right) the RX process. The length of each arrow schematically illustrates the intensity of each variable dependency.}
    \label{fig:diagram}
\end{figure}

Ultimately, models like the present one can be used to tune the microstructural response of W-based materials under fusion operation. Figure \ref{fig:diagram} contains a schematic diagram of the effect of key operation variables on the kinetics of W recrystallization. The direction of the arrows indicates whether a given variable delays (pointing right) or accelerates (pointing left) the RX process as they increase in magnitude. The length of each arrow schematically illustrates the intensity of each variable dependency. A device designer can thus assess how different operational variables will impact the lifetime of key reactor chamber materials and make determinations as to what ranges are considered tolerable for facility operation.

It is worth mentioning, however, that here we have only studied the impact of a single variable at a time (with all others kept unchanged). Thus, we do not discount the existence of cooperative or compounded effects due to the co-interaction between two or more variables. We leave such analyses for future studies.

\section{Summary and conclusions}\label{sec:conc}

In this work, we have developed a coupled framework that simulates the process of [warm/hot working]$\rightarrow$[neutron irradiation]$\rightarrow$[recrystallization] typical of fusion-grade tungsten structures under fusion device operation. 
The model integrates finite-element crystal plasticity simulations of tungsten polycrystals, stochastic cluster dynamics simulations of irradiation in a DEMO fusion environment, and vertex dynamics simulations of microstructure evolution and recrystallization.
Our main concluding points are listed below:
\begin{itemize}
    \item Our work undertakes the important task, essential in modern computational simulation methodologies to have useful and practical applications, of integrating physics models --each tackling a complex aspect of materials evolution-- in a coupled way, with information passing workflows that reflect the physics of the process.
    \item Our simulations reveal that neutron irradiation significantly reduces the effective recrystallization temperature compared to unirradiated microstructures, dropping it from approximately 1400 to 1180 K under DEMO-relevant conditions.
    \item While irradiation initially accelerates recrystallization, the production of Re through neutron transmutation acts as a stabilizer: Re segregation at grain boundaries increases the activation energy for GB motion, effectively restoring or even elevating the recrystallization temperature limit to around 1550 K.
    \item The model predicts rapid irradiation-assisted recrystallization following an initial incubation period, with Avrami exponents consistently near unity, which aligns with experimental observations and points to growth perpendicular to grain boundaries as the operating mechanism during microstructural evolution.
    \item Recrystallization kinetics are exponentially sensitive to temperature due to thermally activated grain-boundary mobility, not vacancy diffusivity. Furthermore, the final recrystallized state exhibits a substantial decrease in flow stress, predicting the typical softening observed in hot-worked and annealed tungsten specimens.
    \item Our model is intended to guide materials design and issue performance predictions under fusion operation conditions. The model permits considering key operational variables in unison: temperature and irradiation push the recrystallization temperature to lower values, while pre-straining and Zener pinning by solid transmutants push it to higher values. 
\end{itemize}
   
\section*{Acknowledgments}

This work was supported by the U.S.~Department of Energy, Office of Science, Office of Advanced Scientific Computing Research and Office of Fusion Energy Sciences, Scientific Discovery through Advanced Computing (SciDAC) program under Award No.~DE-SC0024401. 
Work also partially supported by the Advanced Research Projects Agency - Energy (ARPA-E) under the CHADWICK Program Award Number DE-AR0001993.

\appendix
\section{Model parameters}
The temperature-dependent shear modulus is defined as
\begin{equation}
    \mu(T) = \frac{\sqrt{C_{44}(T)\,[C_{11}(T) - C_{12}(T)]}}{2},
    \label{eq:shear_modulus}
\end{equation}
where the elastic constants $C_{ij}(T)$ are taken from experimental data~\cite{lowrie1967single,cereceda2016unraveling}.

\begin{table}[htbp]
\centering
\caption{Temperature dependent dislocation-dislocation interaction coefficient $h_{\text{dis}}(T)$~\cite{terentyev2015dislocation}.}\label{table:hdis}
\begin{tabular}{|c|c|}
\hline
$T$ [K] & $h_{\text{dis}}(T)$ \\
\hline
600   & 0.14 \\
650   & 0.12 \\
700   & 0.10 \\
750   & 0.08 \\
$> 800$ & 0.06 \\
\hline
\end{tabular}
\end{table}

\begin{table}[htbp]
\centering
\caption{Parameters for tungsten used in the single-crystal plasticity model~\cite{terentyev2015dislocation}.}\label{table:param}
\begin{tabular}{|@{}|l|l|l|@{}|c|}
\hline
Parameter & Definition & Value & Units\\ \hline
$N_s$ & Number of slip systems & 12 & - \\
$\dot{\gamma}_0$ & Reference shear rate & $10^{-3}$ & s$^{-1}$ \\
$m$ & Strain-rate sensitivity exponent & 20 & - \\
$\tau_{\mathrm{f}0}$ & Friction stress at 0~K & 2035 & MPa \\
$k_{\mathrm{B}}$ & Boltzmann constant & $8.61\times10^{-25}$ & eV$\cdot$K$^{-1}$ \\
$\dot{\gamma}_{\mathrm{p}0}$ & Reference strain rate & $3.71\times10^{10}$ & s$^{-1}$ \\
$H_k$ & Kink-pair activation enthalpy & 1.0 & eV \\
$b$ & Burgers vector magnitude & 2.74 & \AA \\
$k_1$ & Kocks–Mecking parameter & $4.8\times10^6$ & m$^{-1}$ \\
$\chi^{\alpha}$ & Interaction parameter & 0.9 & - \\
$g^{\alpha}$ & Normalized activation energy & $2.8\times10^3$ & - \\
$D^{\alpha}$ & Proportionality constant & $10^4$ & MPa \\
$\dot{\varepsilon}_0$ & Reference strain rate & $10^7$ & s$^{-1}$ \\ \hline
\end{tabular}
\end{table}

\begin{table}[htbp]
\centering
\caption{Grain boundary parameters for different misorientations along the [11$\bar{0}$] tilt axis.
The superscript `P' denotes pristine~\cite{mathew2022interstitial}. 
}\label{table:mobility_parameters}
\vspace{10pt}
\begin{tabular}{|c|c|c|c|c|c|c|}
\hline
Tilt Axis & Misorientation [$\circ$)] & $f_0^{\text{P}}$ [s$^{-1}$)] & $Q_a^{\text{P}}$ [eV] & $h$ [\AA] & $b$ [\AA] & $A$ [nm$^2$] \\
\hline
$[11\bar{0}]$ & 26.525 & $4.0 \times 10^{17}$ & 3.457 & 1.58 & 0.8 & 29.20 \\
$[11\bar{0}]$ & 58.992 & $4.9 \times 10^{17}$ & 12.476 & 2.6 & 2.3 & 27.17 \\
$[11\bar{0}]$ & 65.467 & $1.0 \times 10^{16}$ & 8.633 & 2.54 & 2.0 & 30.92 \\
$[11\bar{0}]$ & 70.528 & $1.7 \times 10^{14}$ & 2.701 & 1.3 & 0.9 & 28.94 \\
$[11\bar{0}]$ & 77.884 & $2.0 \times 10^{15}$ & 4.320 & 2.66 & 1.5 & 42.62 \\
$[11\bar{0}]$ & 82.945 & $1.8 \times 10^{16}$ & 1.607 & 2.66 & 1.3 & 25.25 \\
$[11\bar{0}]$ & 86.627 & $2.3 \times 10^{18}$ & 2.037 & 2.66 & 1.0 & 29.32 \\
\hline
\end{tabular}
\end{table}

\begin{table}[htbp]
\centering
\caption{SCD model parameters}
\label{tab:irr}
\renewcommand{\arraystretch}{1.2} 
\setlength{\tabcolsep}{8pt} 
\begin{tabular}{|c|c|c|c|}
\hline
Quantity & Symbol & Value & Units \\
\hline
Neutron spectrum & Fusion & DEMO-hcll & - \\
PKA rate & $\dot\phi$ & $4.59 \times 10^{14}$ & s$^{-1}$ cm$^{-3}$ \\
Dose rate &  & $1.47\times10^{-7}$ & dpa$\cdot \text{s}^{-1}$ \\
Mean PKA energy & $\langle E \rangle$ & 5.9 & keV \\
Representative volume & $\Omega$ & $10^{-18}$ & m$^3$ \\
\hline
\end{tabular}
\end{table}

\bibliographystyle{elsarticle-num}
\bibliography{refs-jaime,refs-from-RXpaper,mybib,cas-refs,Biblio,IFE-refs,refs-he-h,newrefs,refs-from-Hpaper,references,newrefs1}
\end{document}